\documentclass[pdflatex,sn-mathphys-num]{sn-jnl}

\usepackage{graphicx}%
\usepackage{multirow}%
\usepackage{amsmath,amssymb,amsfonts}%
\usepackage{amsthm}%
\usepackage{mathrsfs}%
\usepackage[title]{appendix}%
\usepackage{xcolor}%
\usepackage{textcomp}%
\usepackage{manyfoot}%
\usepackage{booktabs}%
\usepackage{algorithm}%
\usepackage{algorithmicx}%
\usepackage{algpseudocode}%
\usepackage{listings}%
\usepackage{caption}
\usepackage{subcaption}
\usepackage{overpic}
\usepackage{amssymb}
\usepackage[table]{xcolor}  
\usepackage{longtable}
\usepackage{booktabs}
\usepackage{caption}
\usepackage{siunitx}
\usepackage{threeparttable}
\usepackage{tabularx}

\theoremstyle{thmstyleone}%
\theoremstyle{thmstyletwo}%

\theoremstyle{thmstylethree}%

\begin{document}

\title[Article Title]{Learning the PTM Code through a Coarse-to-Fine, Mechanism-Aware Framework}


\author[1]{\fnm{Jingjie} \sur{Zhang}}
\equalcont{These authors contributed equally to this work.}

\author[1]{\fnm{Hanqun} \sur{Cao}}
\equalcont{These authors contributed equally to this work.}

\author[1]{\fnm{Zijun} \sur{Gao}}

\author[3]{\fnm{Yu} \sur{Wang}}

\author[1]{\fnm{Shaoning} \sur{Li}}

\author[4]{\fnm{Jun} \sur{Xu}}

\author[1]{\fnm{Cheng} \sur{Tan}}

\author[5]{\fnm{Jun} \sur{Zhu}}

\author*[2]{\fnm{Chang-Yu} \sur{Hsieh}}\email{kimhsieh@zju.edu.cn}

\author*[1]{\fnm{Chunbin} \sur{Gu}}\email{gchb4science@gmail.com}

\author[1]{\fnm{Pheng Ann} \sur{Heng}}

\affil[1]{\orgdiv{Department of Computer Science and Engineering}, \orgname{The Chinese University of Hong Kong}}

\affil[2]{\orgdiv{College of Pharmaceutical Sciences}, \orgname{Zhejiang University}}

\affil[3]{\orgdiv{School of Computer Science}, \orgname{Peking University}}

\affil[4]{\orgdiv{Institute of Reproductive Medicine, Medical School}, \orgname{Nantong University}}


\affil[5]{\orgdiv{School of Life Sciences}, \orgname{Tsinghua University}}

\abstract{Post-translational modifications (PTMs) form a combinatorial “code” that regulates protein function, yet deciphering this code—linking modified sites to their catalytic enzymes—remains a central unsolved problem in understanding cellular signaling and disease. We introduce COMPASS-PTM, a mechanism-aware, coarse-to-fine learning framework that unifies residue-level PTM profiling with enzyme–substrate assignment. COMPASS-PTM integrates evolutionary representations from protein language models with physicochemical priors and a crosstalk-aware prompting mechanism that explicitly models inter-PTM dependencies. This design allows the model to learn biologically coherent patterns of cooperative and antagonistic modifications while addressing the dual long-tail distribution of PTM data. Across multiple proteome-scale benchmarks, COMPASS-PTM establishes new state-of-the-art performance, including a 122\% relative F1 improvement in multi-label site prediction and a 54\% gain in zero-shot enzyme assignment. Beyond accuracy, the model demonstrates interpretable generalization, recovering canonical kinase motifs and predicting disease-associated PTM rewiring caused by missense variants. By bridging statistical learning with biochemical mechanism, COMPASS-PTM unifies site-level and enzyme-level prediction into a single framework that learns the grammar underlying protein regulation and signaling.}

\keywords{Post-Translational Modifications, Protein Language Model, Enzyme-Substrate Prediction, Crosstalk, Variant Effect Prediction}

\maketitle

\section{Introduction}

Post-translational modifications (PTMs) constitute a dynamic regulatory layer that orchestrates signal integration, spatiotemporal control, and the functional diversification of proteins \cite{mann2003proteomic,walsh2005protein}. This regulation is often achieved through a combinatorial ``code", where specific patterns of modifications—governed by a network of crosstalk—collectively dictate a protein's fate \cite{deribe2010post,hunter2007age,minguez2015ptmcode}. For example, the Tumor Suppressor p53 serves as a paradigm for this principle, its function dictated not by a single modification, but by a dynamic interplay of phosphorylation, acetylation, and ubiquitination. Deciphering this regulatory code is therefore foundational to understanding health and the molecular basis of diseases like cancer and neurodegeneration, where this PTM circuitry is often rewired \cite{walsh2006post,santos2017protein,zhong2023protein,ramazi2021post}.

However, experimentally deciphering this PTM \emph{code} on a proteome-wide scale remains a formidable challenge. While mass spectrometry has advanced PTM discovery, these approaches are often constrained by high costs and sample requirements \cite{aebersold2003mass,virag2020current}. This inherent limitation has motivated the development of computational models to predict PTM sites. Early single-PTM models delivered valuable site-level cues \cite{kiemer2005netacet,luo2019deepphos,wang2020musitedeep,shrestha2024post,wen2025deepmvp} but treated modifications as isolated events, overlooking the combinatorial logic that governs function. Subsequent efforts evolved to multi-PTM settings. Initial \emph{multi-class} designs, which enforce a one-PTM-per-site constraint, represented an important step but were biologically limiting \cite{tan2024metoken}. In contrast, more recent \emph{multi-label} frameworks, which permit the co-localization of multiple PTMs on a single residue, more faithfully reflect the complexity of cellular signaling \cite{yan2023mind,bozkurt2025astraptm}. Yet, even with these advances, three critical limitations persist. \textbf{First}, most methods lack an explicit mechanism to model the interactions among PTM types. So they have difficulty in learning the complex statistical and biochemical dependencies that govern PTM crosstalk, leading to poor performance in ambiguous, co-modified contexts.\cite{minguez2015ptmcode}. \textbf{Second}, prior models have largely failed to address the profound data-centric challenge of a \textbf{dual long-tail} distribution: a severe inter-class imbalance, where the frequencies of different PTM types are highly skewed, and a profound intra-class imbalance, where for any given modification type, unmodified sites vastly outnumber modified ones (Supplementary Fig. 1). \textbf{Third}, the prediction task is typically confined to site profiling (assessing if a site can be modified) rather than progressing to enzyme assignment (who modifies it), thereby omitting the enzyme-substrate specificity that is the fundamental basis of cellular signaling.

\begin{figure}[!htp]
    \centering
\includegraphics[width=1.0\textwidth]{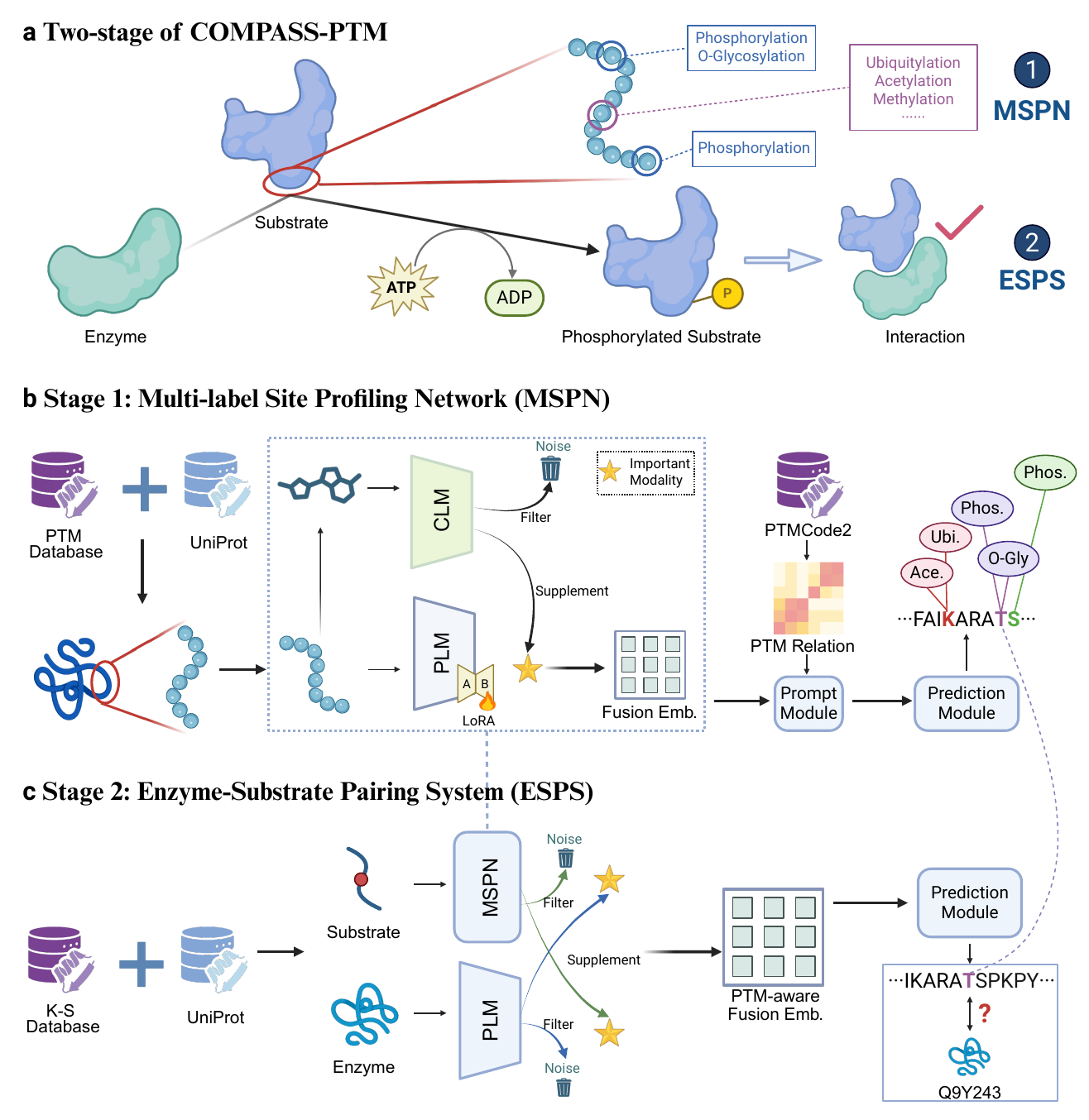}
\caption{\textbf{The COMPASS-PTM framework for mechanism-aware PTM prediction.} \textbf{a}, Conceptual overview of the two-stage, coarse-to-fine framework. Stage 1, the Multi-label Site Profiling Network (MSPN), performs proteome-scale prediction to identify potential PTM sites and their likely modification types. Stage 2, the Enzyme-Substrate Pairing System (ESPS), then takes high-confidence sites from Stage 1 and predicts their cognate enzymes, providing mechanistic context. \textbf{b}, Architecture of the Stage 1 MSPN. A dual-modal encoder fuses representations from a Protein Language Model (PLM), efficiently adapted via LoRA \cite{hu2022lora}, and a Chemical Language Model (CLM). A key innovation, the crosstalk-aware prompting module, uses a PTM relationship matrix—initialized from co-occurrence statistics in the PTMCode2 database \cite{minguez2015ptmcode}—as a learnable inductive bias to guide the prediction of complex, interdependent PTM patterns. \textbf{c}, Architecture of the Stage 2 ESPS. For a given substrate and enzyme pair, the system integrates the PTM-aware substrate embedding generated by the MSPN with a new enzyme embedding to predict their interaction probability, thus linking a modification site to its specific regulator.}
    \label{COMPASS-PTM}
\end{figure}

We address these limitations by reframing PTM inference as a two-stage, \emph{coarse-to-fine}, mechanism-aware learning problem and introduce \textbf{COMPASS-PTM} (Coarse-to-fine Oriented Multi-label PTM site Prediction And enzyme-Substrate aSsignment) (Fig. \ref{COMPASS-PTM}a). 
\textbf{Stage 1} performs proteome-scale, \emph{multi-label site profiling} by fine-tuning a protein language model (PLM) and fusing its evolutionary insights with physicochemical features derived from amino acid properties in a \emph{dual-modal encoder}. Importantly, to explicitly model PTM dependencies, we introduce crosstalk-aware prompting, a component that injects biological priors directly into the learning process. Its innovation lies in a learnable matrix of PTM interdependencies, initialized with crosstalk statistics from PTMCode2 \cite{minguez2015ptmcode}, which steers the model toward biologically plausible co-modification patterns and contextually interdependent predictions, as shown in Fig. \ref{COMPASS-PTM}b. The framework uses a specialized hybrid objective, combining Dice loss for inter-class balance and Focal loss for intra-class balance, to handle the severe dual long-tail data imbalance. \textbf{Stage~2} transitions from site-level profiles to \emph{enzyme–substrate pairing}. For high-confidence sites from Stage 1, a specialized dual-gated residual fusion module integrates local substrate and global enzyme features to assign each site its cognate enzyme, thus translating site predictions into testable mechanistic hypotheses. 

Empirically, the coarse site-profiling phase in Stage 1—empowered by dual-modal integration and crosstalk-aware learning—establishes a new state-of-the-art across three benchmarks developed for this study, delivering up to a 122\% relative improvement in F1-score on the largest and highly imbalanced dbPTM-ML dataset. Building upon this foundation, the fine enzyme–substrate pairing phase in Stage 2 demonstrates robust performance against specialized predictors across diverse benchmarks and split settings. This powerful generalization is further highlighted by its new state-of-the-art performance on the DARKIN zero-shot kinase assignment benchmark \cite{sunar2025darkin}, where it surpasses the previous best method by a remarkable 54\%, confirming the effectiveness of our coarse-to-fine paradigm for generating mechanistically actionable insights.
Beyond quantitative gains, COMPASS-PTM offers qualitative advantages: interpretability grounded in biochemical mechanisms, capacity to predict the functional consequences of pathogenic mutations, and the ability to recover canonical substrate motifs for major enzyme families. These features position COMPASS-PTM as a robust and versatile resource for advancing PTM mechanistic studies and enabling novel biological discovery.

Conceptually, COMPASS-PTM elevates PTM prediction from site classification to \emph{code-aware, mechanism-linked discovery}. Practically, it introduces four key advances:
(i) a crosstalk-aware learner that deciphers the combinatorial PTM code, ensuring predictions reflect the cooperative and antagonistic relationships that govern cellular signaling;
(ii) a dual-modal site representation that integrates deep evolutionary context with local physicochemical principles, creating a more holistic and robust model of substrate suitability;
(iii) a specialized training objective that overcomes the dual long-tail distribution of PTM data;
and (iv) an enzyme-substrate pairing stage that moves beyond site identification to mechanistic insight, linking PTMs to their specific regulators and thus closing the loop from what is modified to who performs the modification.
By tightly coupling statistical learning with biochemical mechanisms, COMPASS-PTM provides a generalizable route toward interpretable machine intelligence. We anticipate this mechanism-aware paradigm will extend beyond PTMs to decode the fundamental regulatory rules of other complex biological systems, such as enzymatic cascades and protein–protein interactions.

\section{Results}\label{results_v2}
\subsection{Overview}
Post-translational modifications (PTMs) represent a fundamental layer of cellular regulation, yet systematic mapping of these modifications and their regulatory enzymes remains a significant challenge. To address this gap, we developed COMPASS-PTM, a deep learning framework that resolves PTM regulatory networks through a coarse-to-fine, mechanism-aware learning approach (Fig. \ref{COMPASS-PTM}a). The framework employs a two-stage architecture that first identifies PTM sites across the proteome and subsequently links these sites to their cognate regulatory enzymes, thereby establishing a comprehensive mapping from modification location to regulatory mechanism.

\paragraph{Stage 1: Multi-label Site Profiling Network}
The first stage, the \textbf{Multi-label Site Profiling Network (MSPN)}, performs proteome-scale PTM site detection (Fig. \ref{COMPASS-PTM}b). It employs a dual-modal fusion architecture integrating evolutionary sequence information from a Protein Language Model (PLM) with physicochemical reactivity determinants from a Chemical Language Model (CLM). Here, PLM features serve as the primary source of information, while CLM features act as auxiliary inputs. The fusion model preserves critical chemical knowledge while filtering out potential noise arising from erroneous predictions in the chemical modality. This multi-dimensional representation simultaneously captures residue conservation patterns and chemical reactivity principles. A key innovation, \emph{crosstalk-aware prompting}, introduces biological interdependence directly into the model's attention mechanism via a learnable prior matrix initialized with empirical PTM co-occurrence statistics \cite{minguez2015ptmcode}. This enables coherent multi-label predictions—for example, increasing confidence in a ubiquitination signal when adjacent phosphorylation is detected—thus echoing known crosstalk mechanisms. To address the dual long-tail distribution inherent in PTM datasets (rare modification types and sparse positive sites), MSPN employs a hybrid loss function (Methods~\ref{subsubsec:foundation-loss}) that prioritizes rare but biologically critical classes while focusing learning on challenging individual instances.

\paragraph{Stage 2: Enzyme-Substrate Pairing System}
The second stage, the \textbf{Enzyme–Substrate Pairing System (ESPS)}, links high-confidence sites from MSPN to their regulatory enzymes. ESPS integrates local substrate sequence motifs with global enzyme embedding representations to capture recognition specificity. In this fusion, both substrate and enzyme features are treated as primary sources of information, retaining key characteristics such as domain architecture while filtering out redundancy and noise from excessively long enzyme sequences. The model directly predicts enzyme–substrate relationships (Fig.~\ref{COMPASS-PTM}c), translating statistical predictions into mechanistic insights. This pipeline produces experimentally testable hypotheses that can elucidate the precise architecture of PTM regulatory networks.

The COMPASS-PTM architecture moves beyond isolated PTM classification by learning a mechanistic map that connects modification sites to their regulatory enzymes. This provides an experimentally actionable framework to systematically identify key regulatory nodes and potential druggable targets within disease-associated signaling pathways.

\subsection{COMPASS-PTM enable comprehensive yet accurate prediction}
\subsubsection{MSPN establishes superior performance across multiple datasets}

\begin{figure}[!htbp] 
    \centering
    \includegraphics[width=1.0\linewidth]{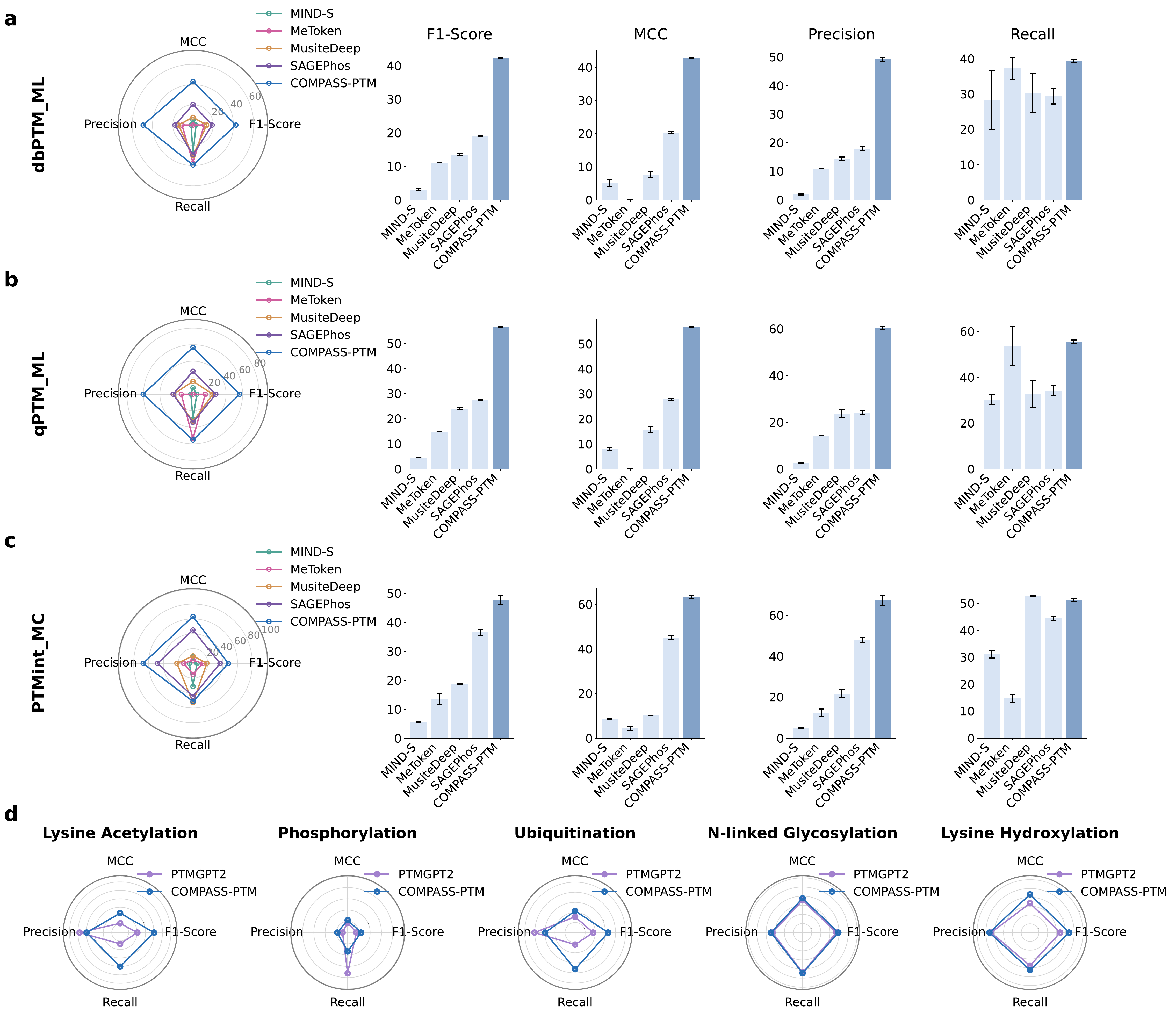}
    \caption{\textbf{MSPN demonstrates state-of-the-art performance and robust generalization across benchmark datasets.} \textbf{a}, Performance comparison with four baselines \cite{yan2023mind, tan2024metoken, wang2020musitedeep, zhang2025sagephos} on the dbPTM-ML multi-label benchmark derived from dbPTM database \cite{chung2025dbptm}. The radar chart displays macro-averaged scores across four key metrics, while the bar plots provide detailed results for each individual metric, with error bars showing standard deviation. \textbf{b}, Performance comparison on the second multi-label benchmark derived from qPTM database \cite{yu2023qptm}, qPTM-ML. \textbf{c}, Zero-shot generalization performance on the PTMint-MC multi-class benchmark derived from PTMint database \cite{hong2023ptmint}, where the model trained on dbPTM-ML was evaluated directly without fine-tuning. \textbf{d}, Cross-task generalization performance on five binary classification datasets from the PTMGPT2 study \cite{shrestha2024post}. For panels \textbf{a}, \textbf{b}, and \textbf{d}, results were averaged across three independent runs with different random seeds. Panel \textbf{c} represents a single, direct inference experiment.}
    \label{fig:stage1}
\end{figure}

To rigorously evaluate the performance and generalization capabilities of the first stage MSPN, we curated a suite of three distinct benchmark datasets from established proteomics resources. Our evaluation suite comprises two multi-label benchmarks, dbPTM-ML and qPTM-ML, which we developed from the dbPTM \cite{chung2025dbptm} and qPTM \cite{yu2023qptm} datasets, respectively, to assess the model on the complex biological reality of sites harboring multiple, co-occurring PTM types. To specifically test the model's generalization to a different task formulation, we also included the PTMint-MC dataset, a multi-class benchmark where each site is annotated with a single PTM type, developed from PTMint dataset \cite{hong2023ptmint}. Detailed specifications for each dataset are provided in the Supplementary Information S.1.1.2. Critically, all three datasets exhibit the dual long-tail distributions characteristic of real-world proteomic data (Supplementary Fig. 1\textbf{b},\textbf{e},\textbf{h},\textbf{j}), providing a stringent and realistic testbed for evaluating proteome-scale prediction models.

We assessed model performance using a suite of metrics (Supplementary Information S.1.2), primarily F1-score and the Matthews Correlation Coefficient (MCC), two metrics renowned for their reliability in evaluating classifiers on imbalanced multi-label data. To rigorously evaluate performance across both common and rare PTMs, all metrics were calculated using a macro-averaging approach that prevents class imbalance from skewing the results.

MSPN demonstrated exceptional performance across both large-scale benchmarks, establishing new performance standards that surpass existing methods. On the dbPTM-ML dataset, as shown in Fig.~\ref{fig:stage1}\textbf{a}, MSPN achieved a macro-averaged F1-score of 0.423, representing a 122\% improvement over the previous best-performing method, SAGEPhos (0.190). This performance gain was accompanied by a corresponding 111\% increase in MCC to 0.429, indicating robust performance across both common and rare modification types.
The qPTM-ML benchmark confirmed these results with similarly improvements, where MSPN again achieved greater than two-fold enhancement in F1-score compared to competing approaches, alongside a 104\% improvement in MCC, as shown in Fig.~\ref{fig:stage1}\textbf{b}. These consistent gains across independent datasets underscore the robustness and generalizability of the MSPN architecture, which effectively encodes rich sequence features and models PTM interdependencies to generate biologically meaningful insights.

A critical distinction emerges when examining the underlying drivers of performance improvement. While several existing methods achieve superficially impressive AUROC scores, such performance often reflects high sensitivity at the expense of prohibitive false-positive rates. For instance, MusiteDeep achieves 0.964 AUROC on dbPTM-ML yet produces a precision of only 0.179, rendering its predictions impractical for experimental validation. In contrast, MSPN's superior F1-score stems from exceptional precision (0.493 on dbPTM-ML), indicating its capacity to generate high-confidence predictions suitable for experimental follow-up.
This high precision is vital for practical utility, as minimizing false positives is essential for guiding limited experimental resources. Therefore, MSPN's balanced performance delivers reliable, experimentally actionable predictions, marking a significant step forward for the field.

The PTMint-MC evaluation provided a stringent test of MSPN's discriminative capability across diverse PTM types and generalization performance, assessing the model's sensitivity to the subtle biochemical signatures that differentiate competing PTM pathways under physiologically relevant but data-constrained scenarios. Despite being optimized for multi-label prediction, MSPN excelled in this multi-class discrimination task, surpassing baseline methods by 30.4\% in F1-score and 40.3\% in MCC, as shown in Fig.~\ref{fig:stage1}\textbf{c}.
MSPN's strong performance on this multi-class task highlights its ability to learn the fundamental, generalizable principles of PTM site recognition.

\subsubsection{MSPN demonstrates robustness in cross-task benchmarks}
To rigorously test the generalizability of our architecture's core principles, we benchmarked MSPN against PTMGPT2, a strong single-PTM predictor, on five of its native binary classification tasks. By retraining our model from scratch for each task—a deliberately challenging setting—we found that MSPN consistently learned more robust representations. This was highlighted by its performance on Lysine Acetylation, where it exceeded the baseline's F1-Score and MCC by 96.8\% and 108.6\%, respectively (see Fig.~\ref{fig:stage1}\textbf{d} and Appendix S.2.1 for full results). This superior cross-task performance demonstrates that our architecture effectively learns the intrinsic rules of PTMs rather than task-specific correlations, validating its robustness as a versatile predictive engine.

\subsubsection{ESPS Achieves Robust and Generalizable Enzyme-Substrate Pairing}

The second stage of our framework, Enzyme-Substrate Pairing System (ESPS), demonstrates exceptional predictive power, establishing new performance benchmarks on both our curated OmniPath dataset \cite{turei2016omnipath} and the established SAGEPhos benchmark \cite{zhang2025sagephos} (see Supplementary Information S.1.1.3 for details). 

\begin{figure}[!htbp]
    \centering
    \includegraphics[width=1.0\linewidth]{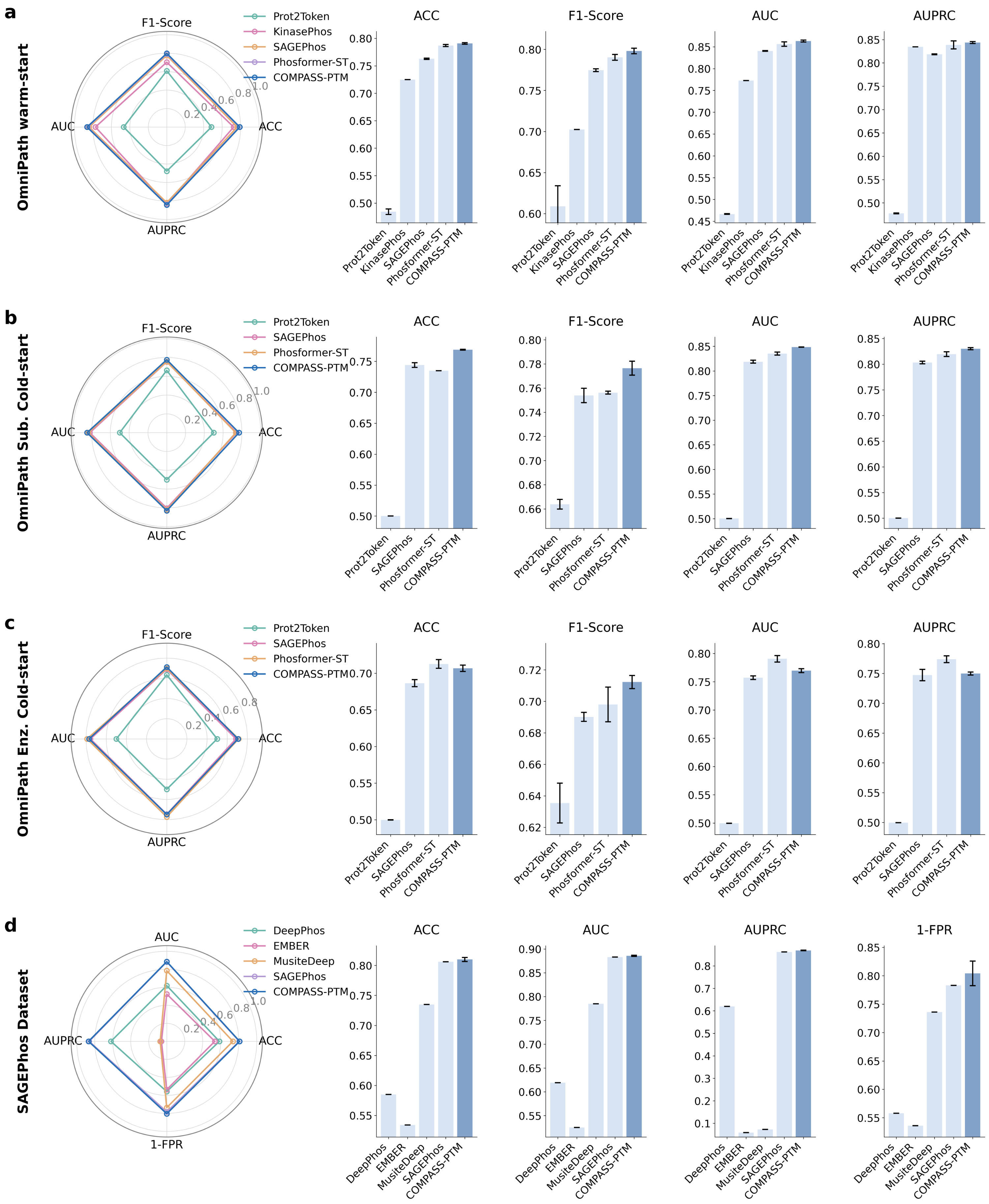}
    \caption{\textbf{ESPS demonstrates robust and high-fidelity performance in enzyme-substrate prediction across diverse benchmarks.} \textbf{a-c}, Performance comparison against competing methods \cite{pourmirzaei2025predicting,ma2023kinasephos,zhang2025sagephos,zhou2024using} on the OmniPath benchmark \cite{turei2016omnipath} across three distinct data splitting strategies: (\textbf{a}) warm-start, (\textbf{b}) substrate cold-start, and (\textbf{c}) enzyme cold-start. The radar chart summarizes performance across four key metrics, while the bar plots show detailed scores for each. \textbf{d}, Independent validation on the SAGEPhos benchmark \cite{zhang2025sagephos}, which uses a distinct suite of metrics including 1-False Positive Rate (1-FPR). All results are the mean of three independent runs with different random seeds; error bars represent standard deviation.}
    \label{fig:stage2}
\end{figure}

On the OmniPath benchmark, ESPS achieved state-of-the-art performance under standard warm-start conditions (random data split), consistently outperforming the leading specialized predictor, Phosformer-ST, with a superior AUC of 86.33\% and AUPRC of 84.36\% (Fig.~\ref{fig:stage2}\textbf{a}). 

To rigorously evaluate the model's generalization capacity, we tested it under two cold-start scenarios, which were defined by partitioning substrates and kinases using MMseqs2. In the substrate cold-start setting—where the model was evaluated on substrate peptides completely absent from the training data—ESPS again outperformed all baseline models across all key metrics (Fig.~\ref{fig:stage2}\textbf{b}), achieving an ACC of 0.769 and an AUC of 0.849. This strong performance underscores the model's ability to capture intrinsic sequence features that determine substrate suitability, a capability we attribute to the discriminative substrate representations learned through the first stage's crosstalk-aware training on multi-label PTM prediction. Under the more challenging kinase cold-start scenario—involving enzymes unseen during training—ESPS remained highly competitive, performing on par with or better than most baselines and only slightly behind Phosformer-ST \cite{zhou2024using} (Fig.~\ref{fig:stage2}\textbf{c}), indicating its robustness in generalizing to novel regulatory contexts.



To further validate these findings, we assessed ESPS on the independent SAGEPhos benchmark under identical experimental conditions (Fig.~\ref{fig:stage2}\textbf{d}). ESPS consistently surpassed the SAGEPhos \cite{zhang2025sagephos} model on its own benchmark across all metrics, including accuracy (80.99\%), AUC-ROC (88.57\%), and AUC-PRC (86.88\%). Notably, it reduced the False Positive Rate to 19.58\%  (compared to 21.7\% by SAGEPhos), a critical improvement for ensuring that experimental validation efforts are directed toward high-confidence candidates. 

It is critical to contextualize this comparison within the broader prediction workflow. While specialized tools achieve high performance on the discrete task of enzyme assignment, they address only the ‘‘fine" inference step, requiring a known site as input. COMPASS-PTM's defining advantage allows it to operate from a full-length protein sequence alone, making it uniquely suited for scenarios where PTM sites are not known a priori.

\subsubsection{Zero-shot Generalization to Unseen Kinases}
Building on our enzyme cold-start experiments that confirmed the model's generalization to unseen enzymes, we further assessed our ESPS's performance on the rigorous DARKIN benchmark \cite{sunar2025darkin}. This benchmark is designed to identify "dark kinases" for a given phosphosite, enforcing strictly disjoint kinase sets between the training and testing splits to ensure a true zero-shot evaluation. Model performance is quantified by the mean Average Precision (mAP), which assesses the ability to correctly rank true kinases for each substrate. For a detailed description of the DARKIN benchmark and the mAP evaluation protocol, see the Supplementary Information S.2.2.

Following the benchmark's protocol, our model achieves the mAP of 0.2946. This result substantially outperforms the highest reported score from the original study (0.1911) by 54\%. Notably, this state-of-the-art performance is achieved using a more parameter-efficient ESM2-150M backbone, in contrast to the larger models previously evaluated. This strong performance underscores our model's robust capability for de-orphanizing kinases and predicting novel enzyme-substrate interactions in a true zero-shot context.


\subsection{Interpretability Analysis Reveals a Learned Biochemical Logic}
To validate that COMPASS-PTM operates not as a ``black box" but as a learner of genuine biochemical principles, we conducted a multi-faceted interpretability analysis. 

\subsubsection{Visualizing a Coarse-to-Fine Representation of PTM Regulation}
We conducted a series of analyses to visualize the learned representations at both stages of the COMPASS-PTM framework. These visualizations reveal that the model self-organizes its embedding space in a hierarchical and biochemically coherent manner, providing a clear mechanistic basis for its predictive power.

\begin{figure}[!htbp]
 \centering
 \includegraphics[width=1.0\linewidth]{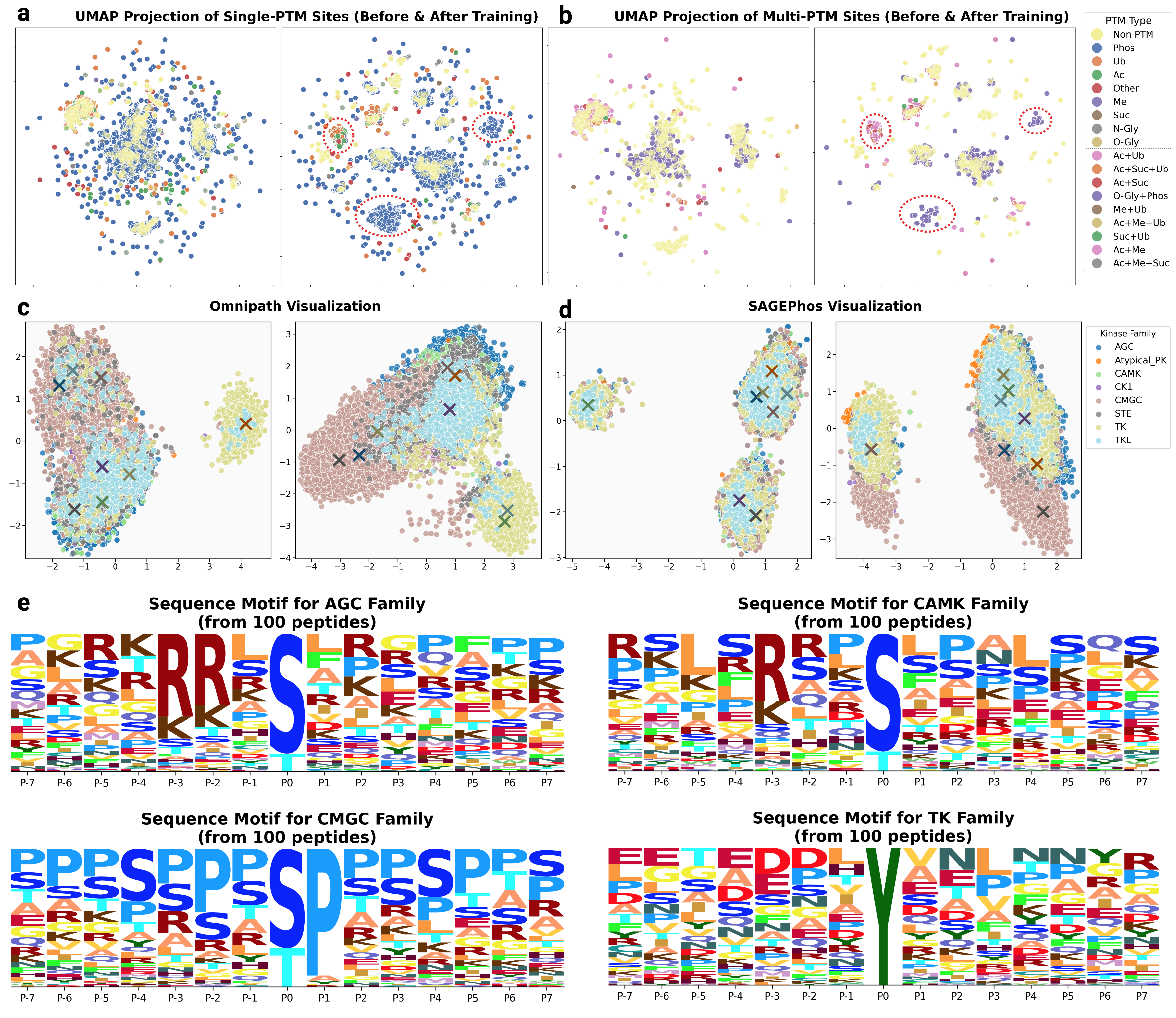}
 \caption{\textbf{Visualizing the learned representations of the COMPASS-PTM framework.} \textbf{a,b}, UMAP projection \cite{mcinnes2018umap} of the Stage 1 embedding space for single-PTM (\textbf{a}) and multi-PTM (\textbf{b}) sites, shown before (left) and after (right) training. \textbf{c,d}, Principal Component Analysis (PCA) \cite{mackiewicz1993principal} of Stage 2 embeddings, organized by kinase family, for the OmniPath (\textbf{c}) and SAGEPhos (\textbf{d}) datasets. For each dataset, clustering based on substrate-only features (left) is coarse, but becomes highly resolved after fusion with enzyme-specific information (right), demonstrating a coarse-to-fine learning dynamic. \textbf{e}, Canonical kinase recognition motifs recovered from the model's top-100 high-confidence predictions for four major kinase families. The model independently rediscovers the well-established biochemical signatures for each family—including the basophilic motifs of AGC and CAMK, the proline-directed motif of CMGC, and the tyrosine-specific motif of TK—validating that its predictions are grounded in the principles of molecular recognition.}
 \label{fig:emb_visual}
\end{figure}

\textbf{Multi-label site profiling demonstrates improved PTM type discrimination through training.}
UMAP (Uniform Manifold Approximation and Projection \cite{mcinnes2018umap}) visualization reveals the progressive, biochemically-aware organization of PTM site representations learned during training. To elucidate this process, we visualized single- and multi-PTM sites separately, holding the projection of non-modified sites constant as a reference frame (Fig. \ref{fig:emb_visual}a-b).
For \textbf{single-PTM sites} (Fig. \ref{fig:emb_visual}a), the initial pre-trained embeddings exist in an undifferentiated state, with PTM and Non-PTM sites heavily intermingled (left panel). Upon training, this space self-organizes into a well-defined and logical structure (right panel). The most fundamental transformation is the segregation of bona fide PTM sites from the Non-PTM background, demonstrating that the model learns to effectively separate biologically meaningful signals from noise. Deeper analysis reveals a sophisticated capacity to resolve fine-grained distinctions that mirror biochemical principles. For instance, the numerous phosphorylation sites resolve into several discrete sub-clusters, strongly suggesting the model has learned to distinguish different phosphorylation contexts, such as motifs recognized by distinct kinase families. The model also deciphers the complex `lysine code': a subset of methylation forms an exclusive cluster, while acetylation and ubiquitination also form their own distinct groups. Although their clusters are in close proximity, they exhibit high internal cohesion, demonstrating that the model learns their unique signatures despite a shared substrate. Furthermore, structurally bulky modifications like N-linked glycosylation converge into a distinct cluster, highlighting the model's ability to learn unique sequence grammars.
A similar organizing principle is evident for \textbf{multi-PTM sites} (Fig. \ref{fig:emb_visual}b). Initially randomly distributed, these sites converge into discrete clusters defined by specific PTM combinations after training. This demonstrates that COMPASS-PTM successfully learns crosstalk relationships, grouping sites that share common combinatorial modification patterns.
In summary, this visualization provides compelling evidence that our model is not a ``black box." Instead, it learns a biochemically coherent map where proximity in the embedding space reflects genuine functional relationships, enabling robust and interpretable predictions.

\textbf{Enzyme-substrate pairing embeddings reveal hierarchical catalytic organization.}
To interrogate the hierarchical representations learned by our two-stage model, we visualized the substrate embeddings from the OmniPath and SAGEPhos datasets using Principal Component Analysis (PCA) \cite{mackiewicz1993principal}, both before and after fusion with enzyme information (Fig. \ref{fig:emb_visual}c-d).
First, we examined the substrate-only embeddings generated by Stage 1 to validate its ability to learn biologically relevant sequence patterns (left panel). The analysis reveals that the embeddings spontaneously organize according to the cognate kinase family that targets them. This emergent organization, a stark contrast to their initial random state (Supplementary Fig. 1\textbf{l-m}), provides compelling evidence that our model successfully extracts the conserved, and often subtle, substrate motifs characteristic of different kinase families. The partial overlap among families like AGC and STE accurately reflects known biological promiscuity in motif recognition.
Next, we visualized the final, fused embeddings from Stage 2, which incorporate information from both the substrate and the specific kinase (right panel). The result is a striking reorganization into discrete, compact, and well-separated clusters, each corresponding to a specific kinase family. This clear segregation provides visual proof that the model has learned the rules of molecular recognition, enabling it to accurately assign substrates to their cognate kinases with high confidence. This progression from general motif learning to specific, accurate interaction prediction was consistently observed across datasets.

\subsubsection{Model Independently Recovers Canonical Kinase Recognition Motifs}
To validate that the predictive accuracy of our Stage 2 model is rooted in a genuine understanding of molecular recognition, we designed an experiment to interrogate its learned representation of enzyme specificity. For each of the eight major kinase families, we selected up to 100 substrate peptides from the test set that COMPASS-PTM predicted with the highest confidence. By generating sequence logos from these high-confidence predictions, we could visualize the sequence features the model deemed most important for each family.

The resulting sequence logos confirmed that COMPASS-PTM robustly recovered the canonical substrate motifs that serve as the biochemical signatures for these kinase families (Fig. \ref{fig:emb_visual}\textbf{e} and Supplementary Fig. 3\textbf{a-d}). For the \textbf{AGC family}, which includes central regulators of cell growth and metabolism such as PKA and Akt, the model correctly identified the classic basophilic motif. This was characterized by a strong preference for basic residues (Arginine/R, Lysine/K) at the P-3 and P-2 positions (where P0 denotes the modified residue and P-n/P+n denote upstream/downstream positions, respectively), conforming to the \emph{R-R-X-S/T} consensus \cite{tegge1995determination}. The model also captured the more complex, hierarchical requirements of the \textbf{CAMK family}, a group responsive to calcium signaling. It identified a stringent preference for Arginine/R at the P-3 position and, critically, a distinct requirement for a large hydrophobic residue (Leucine/L) at the distal P-5 position, consistent with the \emph{L-X-R-X-X-S/T} motif \cite{viengkhou2021novel}. For the \textbf{CMGC family}, comprising essential cell cycle regulators (e.g., CDKs) and signaling kinases (e.g., MAPKs), the model identified the unmistakable proline-directed motif, defined by a near-absolute requirement for Proline/P at both the P-2 and P+1 positions, recapitulating the \emph{P-X-S/T-P} consensus \cite{ryu2009genome,schwartz2005iterative}. Finally, for \textbf{Tyrosine Kinases (TK)}, a family fundamental to growth factor signaling, the model's learned representation extended far beyond the absolute specificity for Tyrosine (Y) at the P-0 site. It correctly captured the strong preference for upstream acidic residues that defines the acidophilic tyrosine motif \cite{ubersax2007mechanisms} and, furthermore, identified a subtle preference for Proline/P at the P+3 position, a feature critical for substrate binding and the assembly of downstream signaling complexes. The independent and accurate recovery of these diverse and nuanced motifs provides compelling evidence that COMPASS-PTM's predictive accuracy is not merely correlational but is instead grounded in a learned understanding of the fundamental, sequence-based rules of enzyme-substrate pairing.

\subsection{Predicting PTM Alterations Driven by Pathogenic Mutations}
To demonstrate COMPASS-PTM's dual utility in both large-scale screening and deep mechanistic inquiry, we applied it to two distinct challenges. First, we performed a systematic screen to identify PTM-disrupting variants in the human sperm proteome. Second, we conducted in-depth case studies on well-characterized disease mutations to validate the model's accuracy and generate novel, precise hypotheses.

\subsubsection{Systematic Screening Identifies PTM-Disrupting Variants in Reproductive Pathology}
To probe the genetic basis of reproductive pathology, a critical yet understudied field, we cross-referenced pathogenic variants from the PhosphoSitePlus PTMVar database \cite{hornbeck2015phosphositeplus} against a comprehensive catalog of the human sperm proteome \cite{castillo2018contribution}. By applying a stringent filtering criterion (wild-type prediction score \textgreater{} 0.8 and mutant prediction score \textless{} 0.2) to pinpoint mutations predicted to abolish PTM events, our screen identified 62 high-confidence SNPs (Supplementary Table 2).
Among these hits, our analysis highlighted several compelling cases directly linked to reproductive pathologies. A notable example is the \textbf{S2621C} variant in the Adenomatous Polyposis Coli (APC) protein, a mutation associated with Familial Adenomatous Polyposis 1 (FAP1). While primarily a cancer syndrome, APC is also vital for normal spermatogenesis \cite{li2016point}. COMPASS-PTM predicted that this mutation causes a catastrophic loss of a key phosphorylation site, with the prediction score plummeting from 0.944 in the wild-type to a negligible 0.0002 in the mutant. Another significant finding emerged for SperMine Synthase (SMS), where the \textbf{K4295N} mutation is linked to an intellectual developmental disorder that often presents with hypogonadism \cite{shangguan2024study}. Our model revealed that this variant likely abolishes a critical ubiquitination site, with the score dropping from a confident 0.869 to just 0.0003. These examples from our screen underscore the power of COMPASS-PTM to generate specific, testable hypotheses linking genetic variants to disease phenotypes by disrupting PTM-mediated regulation.

\subsubsection{Validating Accuracy and Uncovering Disease Pathways via Case Studies}
Beyond large-scale screening, we sought to rigorously assess the model's accuracy on an individual level. We first validated its ability to computationally recapitulate known pathogenic pathways and then deployed it to generate novel hypotheses for uncharacterized variants.

\begin{figure}[!htbp]
    \centering
    \includegraphics[width=1.0\linewidth]{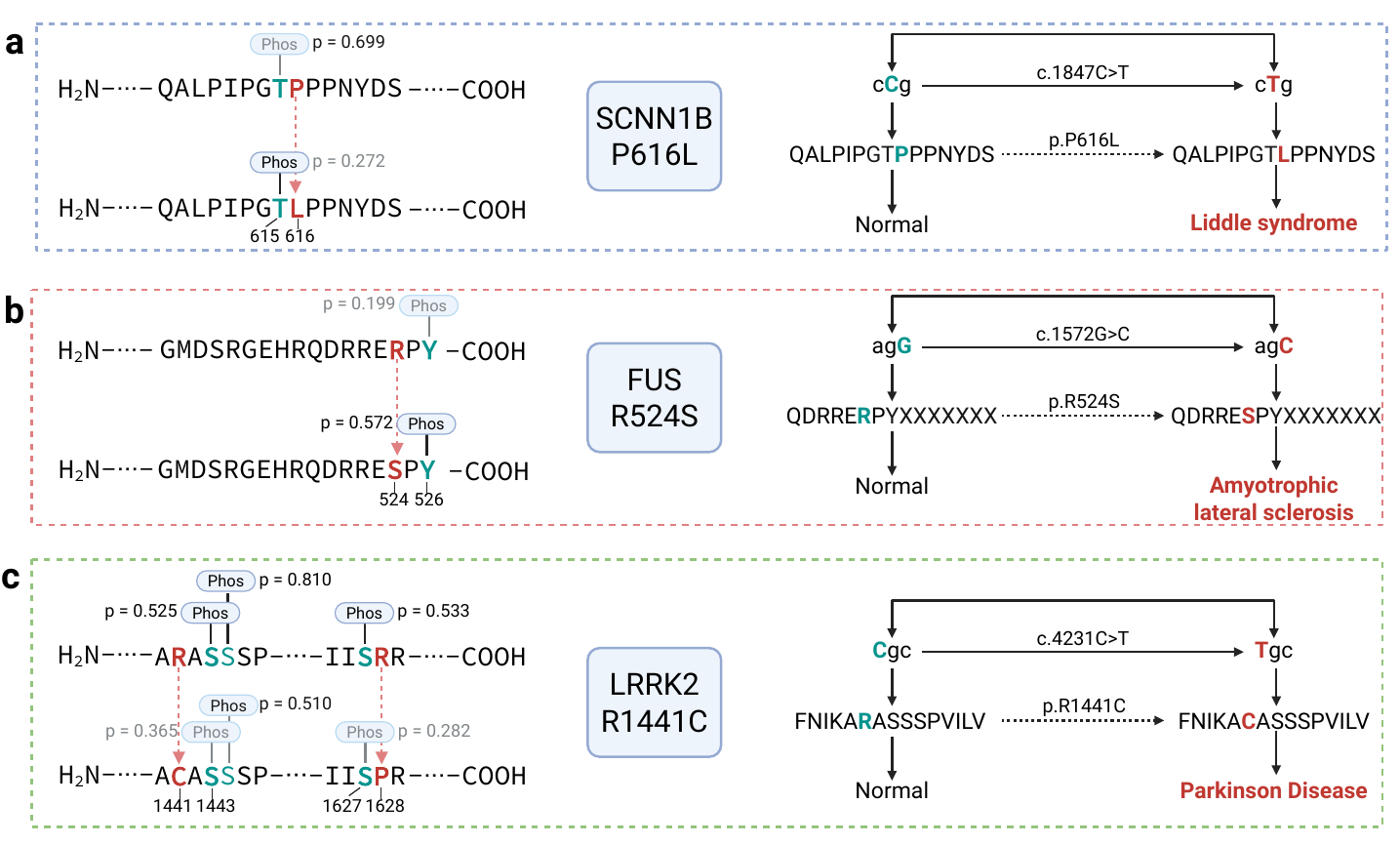}
    \caption{\textbf{COMPASS-PTM generates mechanistic hypotheses by predicting the PTM consequences of pathogenic variants.} \textbf{a}, The Liddle syndrome-associated p.P616L substitution in SCNN1B is predicted to cause a loss of phosphorylation at the adjacent T615 site (predicted probability drops from 0.70 to 0.27) \cite{gwozdzinska2017hypercapnia}. \textbf{b}, The amyotrophic lateral sclerosis (ALS)-associated p.R524S variant in FUS is predicted to induce a gain of phosphorylation at the nearby Y526 site (probability increases from 0.20 to 0.57) \cite{zhang2012structural,darovic2015phosphorylation}. \textbf{c}, Two Parkinson's disease-associated variants in LRRK2, p.R1441C and p.R1628P, are both predicted to result in a loss of phosphorylation at their respective proximal sites \cite{muda2014parkinson}. For each case, the diagrams on the left illustrate the local sequence and predicted PTM probabilities, where the pathogenic substitution is highlighted in red and the affected PTM site is in green. The diagrams on the right summarize the underlying genetic alteration and resulting pathology.}
    \label{fig:mutation}
\end{figure}

We began by examining two well-studied mutations in the LRRK2 gene, a key locus in Parkinson’s disease. Confirming its accuracy, our model predicted that the R1441C variant leads to a loss of phosphorylation at S1443 (Fig. \ref{fig:mutation}c). This result perfectly aligns with extensive experimental evidence that this mutation impairs PKA-mediated phosphorylation, thereby disrupting LRRK2’s interaction with 14-3-3 regulatory proteins—a cornerstone of its pathogenic mechanism \cite{muda2014parkinson}. The model likewise correctly identified that the R1628P variant causes a loss of phosphorylation at S1627, further validating its ability to pinpoint PTM-altering events in a complex disease protein.

Having established the model's accuracy, we then deployed it as a hypothesis-generation engine. For the P616L variant in SCNN1B, which causes Liddle syndrome, the model proposes a precise molecular basis for the known overactivation of the ENaC channel. It predicts a profound loss of phosphorylation at T615 (Fig. \ref{fig:mutation}a), a site whose modification is the known signal for the channel's removal from the cell surface \cite{gwozdzinska2017hypercapnia}. In another case, the model generated a novel hypothesis for the R524S variant in the FUS protein, which is associated with ALS. It predicted a gain of phosphorylation at Y526 (Fig. \ref{fig:mutation}b), suggesting a synergistic pathogenic mechanism where the mutation enhances a PTM event that is also known to impair FUS's nuclear import, thereby exacerbating its toxic cytoplasmic aggregation \cite{zhang2012structural,darovic2015phosphorylation}. See the Apendix S.2.4 for more case studies.

Collectively, these applications demonstrate COMPASS-PTM's versatility as an engine for translational research. It can efficiently prioritize candidates from large-scale genetic data while also providing the granular, mechanistic insights needed to design focused experiments, bridging the gap between genomic variation and disease pathology.

\subsection{Case Studies}

\begin{figure}[!htbp]
    \centering
    \includegraphics[width=1.0\linewidth]{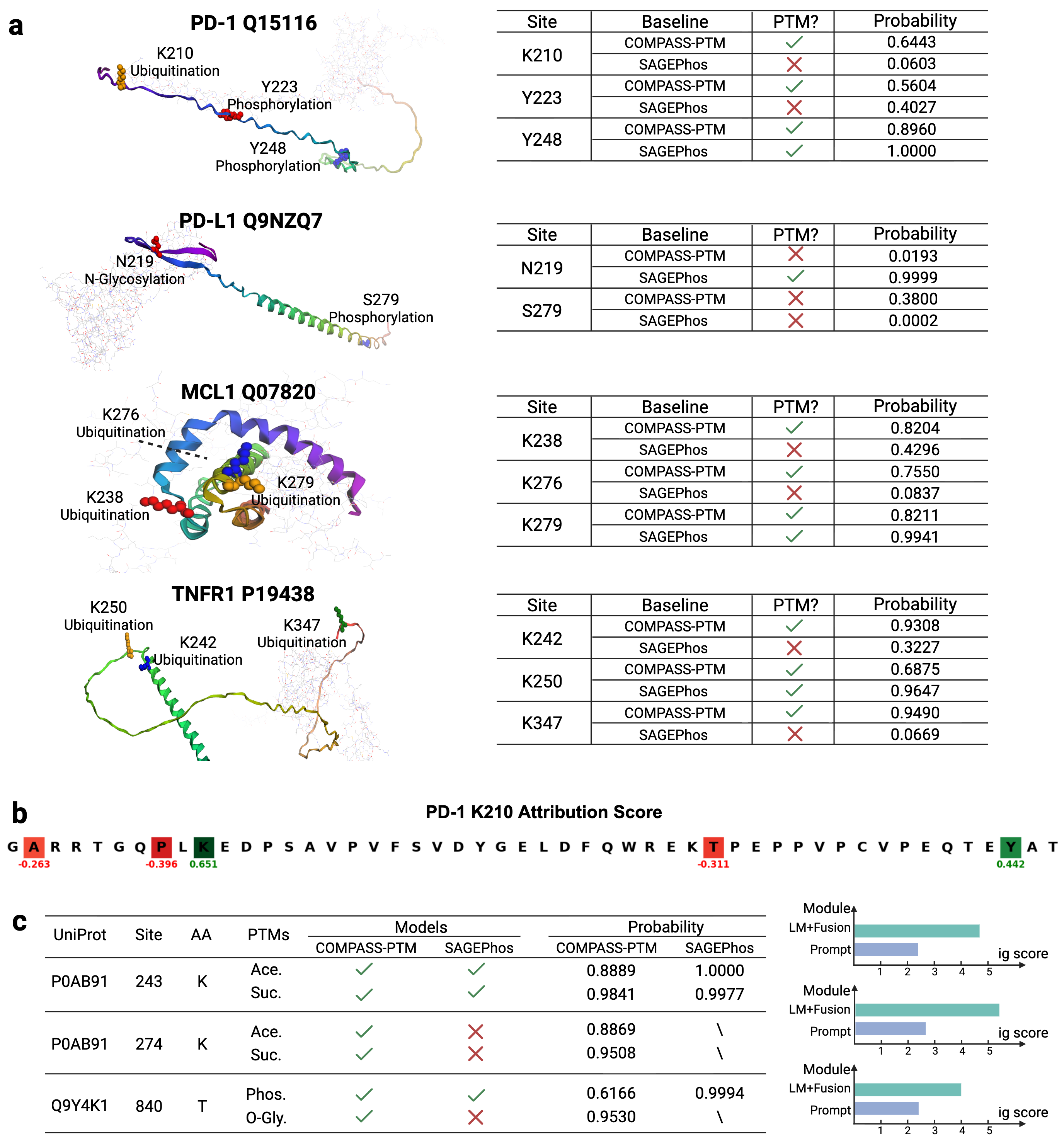}
    \caption{\textbf{Case studies demonstrating the accuracy and mechanistic interpretability of COMPASS-PTM on clinically relevant targets.} \textbf{a}, Zero-shot prediction of functionally critical PTMs on four disease-associated proteins. The tables compare the predictions of COMPASS-PTM against the baseline (SAGEPhos \cite{zhang2025sagephos}) for sites that were withheld from the training set. Correct predictions are marked with a green check and incorrect predictions with a red cross. \textbf{b}, Gradient-based attribution analysis for the prediction of ubiquitination at K210 on PD-1. The height and color of each residue indicate its contribution to the final prediction score (green for positive, red for negative), revealing the sequence determinants the model learned. \textbf{c}, Superior performance in complex multi-label prediction on sites with co-occurring PTMs. The bar plots on the right show attribution scores for the model's core components, quantifying the significant contribution of the crosstalk-aware Prompt module to the correct classification of these sites.}
    \label{fig:case_study}
\end{figure}

\subsubsection{COMPASS-PTM excels at identifying critical regulatory sites}
To assess COMPASS-PTM's capacity for novel discovery on clinically relevant targets, we performed zero-shot predictions on four proteins central to immune regulation and cell survival: PD-1, PD-L1, MCL1, and TNFR1 (Fig. \ref{fig:case_study}a). By benchmarking against the top baseline, SAGEPhos \cite{zhang2025sagephos}, on PTM sites not present in the training set, we evaluated the model's ability to identify functionally critical regulatory sites.

\textbf{PD-1 reveals superior detection of immune checkpoint regulatory modifications.} 
On PD-1, a key immune checkpoint receptor, COMPASS-PTM successfully identified pivotal regulatory modifications missed by the baseline \cite{lee2024postPD-1}. It correctly predicted ubiquitination at K210 (score: 0.6443 vs. 0.0603), a PTM that terminates the inhibitory signal by triggering receptor degradation \cite{meng2018fbxo38}. Furthermore, it accurately identified phosphorylation at Y248 (score: 0.8960), the crucial event for recruiting SHP-2 phosphatases to execute T-cell inhibition \cite{chemnitz2004shp}. This ability to pinpoint distinct molecular switches underscores the model's precision in deciphering immune signaling.

\textbf{PD-L1 modifications reveal complementary predictive capabilities across PTM types.} 
The analysis of PD-L1 revealed complementary strengths between the models, highlighting the complexity of the PTM code \cite{feng2023regulationPDL1}. While SAGEPhos showed high-confidence prediction of N-glycosylation at N219 (0.9999 vs. 0.0193), a PTM that stabilizes the protein, COMPASS-PTM uniquely identified phosphorylation at S279 (0.3800 vs. 0.0002), a site completely missed by the baseline. This demonstrates how different architectures may capture distinct biochemical principles.

\textbf{MCL1 ubiquitination patterns demonstrate superior detection of apoptosis regulatory sites.} 
For the anti-apoptotic protein MCL1, whose stability is governed by ubiquitination-mediated degradation, COMPASS-PTM showed superior detection of key ``degron" sites \cite{wu2020ubiquitinationMCL1}. It confidently identified critical turnover sites at K238 (0.8204 vs. 0.4296) and K276 (0.7550 vs. 0.0837), sites where the baseline was either less confident or failed to make a positive prediction. This robust performance on functionally validated degrons highlights its potential to identify key regulatory switches in cell survival pathways.

\textbf{TNFR1 ubiquitination demonstrates comprehensive regulatory site detection.} 
On TNFR1, a receptor that governs inflammatory and cell death pathways, COMPASS-PTM excelled at identifying ubiquitination sites critical for signaling complex assembly \cite{li2025exploringTNFR1}. It demonstrated markedly superior performance in detecting ubiquitination at K242 (0.9308 vs. 0.3227) and K347 (0.9490 vs. 0.0669), residues essential for recruiting adaptors to the pro-survival ``Complex I," showcasing its ability to predict the assembly points for critical cell fate decisions.

\textbf{Interpretability analysis reveals mechanistic insights into PTM Code.} 
To dissect the sequence determinants underpinning its predictive engine, we employed a robust gradient-based attribution analysis focused on the ubiquitination of lysine 210 (K210) in PD-1 (Fig. \ref{fig:case_study}b). The analysis revealed a remarkable finding: the phosphorylation site at tyrosine 248 (Y248)—the principal switch for PD-1's inhibitory function—emerged as a dominant predictive feature for ubiquitination at the distal K210 site. This provides direct, mechanistic evidence that COMPASS-PTM has learned the principles of PTM crosstalk, as it correctly infers the synergistic relationship where phosphorylation can signal for subsequent ubiquitination, a mechanism often mediated by phospho-degron motifs \cite{ravid2008diversity}. This ability to integrate signals from a distributed network of local, distal, and key functional residues confirms that COMPASS-PTM transcends simple motif-matching, leveraging its dual-modal architecture to achieve its high accuracy.

\subsubsection{COMPASS-PTM Excels in Multi-Label Prediction by Learning Biochemical Constraints}
A critical test for any PTM predictor is its ability to accurately profile sites harboring multiple, concurrent modifications. We benchmarked COMPASS-PTM against the top-performing baseline, SAGEPhos, and found that our model not only achieves higher accuracy but, critically, generates more biochemically coherent predictions (Fig.\ref{fig:case_study}c).
For instance, on protein P0AB91, COMPASS-PTM correctly identifies the co-occurrence of acetylation and succinylation on two distinct lysine residues (K243 and K274), whereas SAGEPhos fails at the latter site, indicating a sensitivity gap. More strikingly, on a threonine residue (T840) of protein Q9Y4K1, COMPASS-PTM successfully resolves the complex crosstalk between phosphorylation and O-linked Glycosylation. SAGEPhos, while correctly identifying the phosphorylation event with high confidence, completely fails to detect the concurrent O-linked Glycosylation.

A deeper analysis of the SAGEPhos predictions for T840 reveals a critical flaw in its learning paradigm. While the baseline model can assign high probabilities to some correct predictions, this appears to be a symptom of a broader tendency to generate spurious, high-probability outputs. On the threonine T840 site, SAGEPhos not only correctly identifies phosphorylation but also erroneously predicts two lysine-exclusive modifications—acetylation and ubiquitination—with biochemically impossible and misleadingly high probabilities.

In contrast, COMPASS-PTM's ability to consistently generate plausible predictions proves it has learned these foundational rules. The source of this superior biochemical reasoning is quantifiable: a gradient-based attribution analysis shows our crosstalk-aware prompting mechanism—the module explicitly modeling PTM dependencies—consistently accounts for approximately one-third of the predictive power in these complex cases (Fig.\ref{fig:case_study}c, right panels).

\subsection{Systematic Ablation Studies Validate the COMPASS-PTM Architecture}
To systematically validate our design principles, we conducted a series of ablation experiments to interrogate the contribution of each component in the COMPASS-PTM framework. For the Stage 1 Multi-label Site Profiling Network (MSPN), we first confirmed our empirical selection of the optimal backbones—ESM2-150M for the protein language model and SELF-BART for the chemical language model—and the superiority of our proposed dual-modal fusion architecture over simpler methods. More critically, these studies quantified the substantial impact of our key innovations. The crosstalk-aware prompting module proved essential for modeling PTM interdependencies, as its removal resulted in a 10.7\% decrease in F1-score and a 9.8\% decrease in MCC. Parameter-efficient fine-tuning with LoRA was also indispensable, boosting overall performance by approximately 30\% and precision by 44\%, which underscores the necessity of adapting the large PLM to the specific task of PTM profiling.
Finally, our hybrid loss function was confirmed to be highly effective, successfully balancing the precision-recall trade-off to outperform standard loss functions in the context of the severe dual long-tail data distribution (see Supplementary Note S.2.6.1 and Supplementary Fig. 5 a-f for full details).

The design of the Stage 2 Enzyme-Substrate Pairing System (ESPS) was similarly validated. Our proposed dual-gated residual fusion module demonstrated superior performance against standard fusion strategies, including concatenation and cross-attention, confirming its effectiveness in selectively integrating the most informative features from both the enzyme and substrate. Furthermore, we determined that a 15-amino-acid peptide length provided the optimal local context for the pairing task, yielding higher performance than both shorter and longer sequences. Together, these analyses provide a clear rationale for our methodological choices and quantify the synergistic contributions of each component to the framework's overall success (see Supplementary Note S.2.6.2 and Supplementary Fig. 5 g-h for full details).

\section{Discussion}\label{discussion}
This work reframes post-translational modification (PTM) prediction as a mechanism-aware learning challenge, using biologically grounded inductive biases to connect statistical prediction with biochemical reality. COMPASS-PTM integrates site-level and enzyme-level prediction within a single, coarse-to-fine architecture, yielding interpretable outputs that link modification patterns directly to their regulatory enzymes.

Beyond establishing new performance benchmarks, our framework demonstrates how incorporating structured biological priors—specifically PTM crosstalk relationships—can guide deep models toward meaningful mechanistic insights. The model's capacity to recover established kinase motifs and accurately predict the effects of pathogenic variants suggests a path for machine learning to evolve beyond mere pattern recognition toward genuine mechanistic discovery.

Current limitations largely mirror field-wide challenges: database biases, incomplete labeling, and representations that prioritize sequence over 3D structure and spatiotemporal context. Addressing these will require developing uncertainty-aware modeling, integrating structure-based representations, and creating benchmarks that accurately reflect biological complexity and imbalance.

We anticipate this mechanism-aware approach will extend beyond PTMs to other biochemical processes, such as complex enzymatic cascades and protein–protein interactions. By tightly coupling statistical learning with mechanistic reasoning, frameworks like COMPASS-PTM may offer a general route toward interpretable machine intelligence for decoding the fundamental regulatory rules of life.

\section{Methods}
\subsection{Problem Formulation}
\label{subsec:problem}
We addresses two coupled prediction tasks concerning post-translational modifications (PTMs).

\textbf{Task1: Multi-label PTM Site Profiling.}
The primary task is formalized as a residue-wise multi-label classification problem. Given a substrate protein sequence $S = \{s_i\}_{i=1}^L$, the objective is to predict a $C$-dimensional binary label vector $\mathbf{y}_i \in \{0,1\}^C$ for each residue $s_i$, where $C$ is the number of PTM types considered for a given task. Each dimension in this vector corresponds to a specific modification class (e.g., phosphorylation, acetylation) known to target distinct amino acid residues. The prediction function $f_\theta$ thus maps a sequence and a position to a probabilistic output for each modification type $c$:
\begin{equation}
    \mathbf{y}_i^c = f_\theta(S,i,c), \quad \forall c \in \{1, \dots, C\}
\end{equation}
where $\mathbb{P}(y_i^c=1|S) \in [0,1]$ is the predicted probability for the residue-type pair $(s_i, c)$.

\textbf{Task2: Enzyme-Substrate Pairing.}
The second, more fine-grained task is to identify the specific enzyme responsible for a modification. Given a substrate sequence $S$ and an enzyme sequence $E = \{e_j\}_{j=1}^M$, the prediction function $g_\phi$ aims to identify the specific site of modification:
\begin{equation}
    \mathbf{z}_i^{c,E} = g_\phi(S,E,i,c), \quad \forall c \in \{1, \dots, C\}, \forall E \in \mathcal{E}_c
\end{equation}
where $\mathbb{P}(z_i^{c,E}=1|S,E) \in [0,1]$ represents the probability that enzyme $E$ from the set of valid catalysts $\mathcal{E}_c$ performs a type $c$ modification at residue $s_i$.

We developed COMPASS-PTM, a unified coarse-to-fine two-stage computational framework, to address these tasks. The first stage, a Multi-label Site Profiling Network (MSPN), performs Task 1. The second stage, an Enzyme-Substrate Pairing System (ESPS), leverages the outputs of the first stage to perform Task 2.

\subsection{Stage 1: Multi-label crosstalk-aware Site Profiling Network}
\label{subsec:foundation}
The first stage of our framework is designed to predict multi-label PTM sites while explicitly modeling the interdependencies between different modification types.
\subsubsection{Dual-Modal Feature Representation} 
\label{subsubsec:feature_encoding}

Our framework employs dual-branch encoding to derive a comprehensive representation for substrate peptides $S_{\text{peptide}}$, integrating complementary large-scale biological context and fine-grained physicochemical properties.

\textbf{Biological Context Modality:} 
We leverage the evolutionary-scale knowledge within a protein language model (PLM) ESM2-150M \cite{lin2023evolutionary} to extract context-aware features $\mathbf{H}_s^{\text{prot}} \in \mathbb{R}^{d_p}$ ($d_p$ = 640). To specialize the PLM for PTM substrate recognition, we implement Low-Rank Adaptation (LoRA) \cite{hu2022lora}, a parameter-efficient fine-tuning technique. 
This adaptation enhances detection of evolutionarily conserved motifs and structural propensities pertinent to modification sites.

\textbf{Chemical Attribute Encoding:}
To capture atomic-level properties, we generate a unified chemical representation $\mathbf{H}_s^{\text{chem}}$ by concatenating learned embeddings from SELF-BART \cite{priyadarsini2024selfbart} with explicit per-residue physicochemical descriptors $\mathbf{P}_s$ (e.g. hydrophobicity, polarity):


\subsubsection{Bio-Coupled and Augmented Fusion}
\label{subsubsec:fusion}
To effectively integrate the rich, evolutionarily-informed features from PLMs ($\mathbf{H}_s^{\text{prot}}$) with potentially noisy, residue-based chemical attributes ($\mathbf{H}_s^{\text{chem}}$), we developed a Bio-Coupled and Augmented Fusion module. This approach establishes a clear information hierarchy: the robust biological representation serves as the primary modality, which is then selectively refined and augmented by the auxiliary chemical information.

The fusion process is governed by a dynamic mechanism that additively combines a gated component ($\mathbf{\Phi}$) and a residual component ($\mathbf{\Psi}$), controlled by learnable parameters $\alpha$ and $\beta$: $\mathbf{H}_s^{\text{fused}} = \alpha \cdot \mathbf{\Phi}(\mathbf{H}_s^{\text{prot}}, \mathbf{H}_s^{\text{chem}}) + \beta \cdot \mathbf{\Psi}(\mathbf{H}_s^{\text{prot}}, \mathbf{H}_s^{\text{chem}})$.
Here, the gating mechanism $\mathbf{\Phi}$ learns to filter the chemical features, preserving the primary protein signal while allowing only the most task-relevant chemical information to modulate it. Concurrently, the residual mechanism $\mathbf{\Psi}$ is designed to capture valuable cross-modal interactions, enriching the primary feature space with complementary chemical context. The details of both the gating and residual mechanisms are provided in the Supplementary Methods S.3.1.1.


\subsubsection{Crosstalk-Aware Prompting PTM Attention}
\label{subsubsec:crosstalk}
A key innovation of our model lies in conceptualizing PTM interdependence as a crosstalk-aware prompting mechanism. To model the biological principle that PTMs do not occur in isolation, we designed a custom attention mechanism that incorporates PTM interdependencies as an inductive bias. This is achieved by injecting a dynamic prompt into the self-attention calculation.

The foundation of this prompt is a learnable $C \times C $ matrix $\mathbf{R}$, which is initialized with a prior based on known PTM co-occurrence statistics ($\mathbf{R}_{\text{prior}}$) and fine-tuned during training to capture complex, non-linear dependencies. The details of $\mathbf{R}$ and $\mathbf{R}_{\text{prior}}$ are in Supplementary S.3.1.2.

Crucially, this learned matrix is used to generate a dynamic prompt tailored to the specific context of each residue. For each position $i$ in a sequence, the model first generates a preliminary PTM probability distribution, $\mathbf{p}_i = \text{softmax}(\text{PTM-Predictor}(\mathbf{H}_i)$, from its final hidden-state embedding $\mathbf{H}_i$.
Using these probabilities, we dynamically compute a site-pair-specific attention prompt, $\mathbf{B}$. The prompt value between any two positions, $i$ and $j$, quantifies the biochemical plausibility of their co-modification, mediated by the learned relationship matrix $\mathbf{R}$. This interaction term is passed through a `tanh' activation, regularized with `Dropout', and scaled by a learnable parameter $\alpha$:

\begin{equation}
    \mathbf{B}_{ij} = \text{Dropout}(\tanh(\mathbf{p}_i^T \mathbf{R} \mathbf{p}_j)) \cdot \alpha
\end{equation}

The resulting bias matrix $\mathbf{B}$ is added directly to the attention logits, thus explicitly guiding the model to focus on residue pairs with high predicted biochemical synergy:
\begin{equation}
    \text{Attention}(\mathbf{Q}, \mathbf{K}, \mathbf{V}) = \text{softmax}\left(\frac{\mathbf{Q}\mathbf{K}^T}{\sqrt{d_k}} + \mathbf{B}\right)\mathbf{V}
\end{equation}
By compelling the model to reason about PTMs as an interconnected system, this crosstalk-aware prompting mechanism ensures that the output is not merely a set of independent site predictions, but a single, biologically coherent combinatorial profile.

\subsubsection{Balanced Training Objective}
\label{subsubsec:foundation-loss}
To address the severe data imbalance inherent in PTM prediction—characterized by both a long-tail distribution across PTM types (inter-class imbalance) and a vast excess of unmodified sites (intra-class imbalance)—we designed a hybrid loss function. This objective synergistically combines a macro-level regularized Dice loss $\mathcal{L}_{\text{macro}}$ with a micro-level Focal loss $\mathcal{L}_{\text{micro}}$, with their contributions balanced by a learnable parameter, $\eta$:

\begin{equation}
    \mathcal{L}_{\text{stage1}} = \eta \mathcal{L}_{\text{macro}} + (1-\eta) \mathcal{L}_{\text{micro}}
    \label{loss_equ}
\end{equation}

The macro-level component, $\mathcal{L}_{\text{macro}}$, is designed to tackle inter-class imbalance. It is composed of a Dice loss, $\mathcal{L}_{\text{Dice}}$, which ensures that rare PTM types contribute equally to the training objective, and a custom Magnification loss, $\mathcal{L}_{\text{Mag}}$, weighted by a learnable parameter $\omega$:
\begin{equation}
    \mathcal{L}_{\text{macro}} = \mathcal{L}_{\text{Dice}} + \omega \mathcal{L}_{\text{Mag}}
\end{equation}
The Magnification loss acts as a direct regularizer on the output probabilities to mitigate over-confidence and improve model calibration, which is particularly crucial for sparse data: $\mathcal{L}_{\text{Mag}}=\mathbb{E}[\sigma(\hat{\mathbf{y}})]$,
where $\sigma$ represents the sigmoid activation function, which transforms the model's raw output logits, $\hat{\mathbf{y}}$, into class-wise probabilities. 

The micro-level component, $\mathcal{L}_{\text{micro}}$, is the Focal Loss, which addresses the intra-class imbalance between the few modified sites and the vast number of unmodified ones.
The detailed mathematical formulations for both $\mathcal{L}_{\text{Dice}}$ and 
$\mathcal{L}_{\text{micro}}$ are provided in the Supplementary Methods S.3.1.3. 

\subsection{Stage 2: Enzyme-Substrate Pairing System}
\label{subsec:enzyme-integration}
The second stage predicts enzyme-substrate specificity by integrating features from both the enzyme and the substrate. We freeze the weights of the Stage 1 network and use it to generate a local substrate representation, $\mathbf{H}{\text{substrate}}$, for a 15-mer peptide centered on the candidate site. The global enzyme representation, $\mathbf{H}{\text{enzyme}}$, is derived from its full-length sequence using ESM-2 \cite{lin2023evolutionary}.
To model the binding interaction, we developed a Dual-Gated Residual Fusion module.This architecture concurrently preserves essential information from both inputs while capturing emergent interaction features. It employs two independent gating mechanisms ($\mathbf{g}_{\text{substrate}}$, $\mathbf{g}_{\text{enzyme}}$) to selectively retain salient features from the original inputs, while a parallel residual network captures emergent binding characteristics ($\mathbf{H}_{\text{residual}}$). These three components are integrated via a learnable weighted sum:
\begin{equation}
    \mathbf{H}_{\text{fused}} = \alpha_0 (\mathbf{g}_{\text{substrate}} \odot \mathbf{H}_{\text{substrate}}) + \alpha_1 (\mathbf{g}_{\text{enzyme}} \odot \mathbf{H}_{\text{enzyme}}) + \alpha_2 \mathbf{H}_{\text{residual}}
\end{equation}
The learnable scalars ($\alpha_i$) enable the model to dynamically balance the contributions of substrate motifs and global enzyme features, enhancing prediction robustness. Further implementation details are provided in Supplementary Methods S.3.2.1.

\section*{Data availability}
    This study utilizes five datasets focusing on two stage predictions. For the first stage (Multi-label Site Profiling), three raw Post-Translational Modifications (PTMs) datasets were used: dbPTM, accessible from \url{https://biomics.lab.nycu.edu.tw/dbPTM/download.php}; qPTM, available from \url{https://qptm.omicsbio.info/download.php}; and PTMint, obtainable from \url{https://ptmint.sjtu.edu.cn/Download}. For the second stage (Enzyme–Substrate Pairing), we used two enzyme-PTM relationships datasets: OmniPath, accessible from \url{https://omnipathdb.org/}; and SAGEPhos dataset, available from \url{https://github.com/ZhangJJ26/SAGEPhos/releases}. All of the Full-length protein sequences are extracted from UniProt \url{https://www.uniprot.org/}.

\section*{Code availability}
The COMPASS-PTM framework utilizes several open-source tools and packages that are integral to its functionality. These tools are publicly available and can be accessed through their respective GitHub repositories or websites.

For protein language model, we utilized ESM2, which is available at \url{https://github.com/facebookresearch/esm}.
For the LoRA fine-tuning process, COMPASS-PTM employs the PEFT package, available at \url{https://github.com/huggingface/peft}.
Sequence similarity searches are performed using MMseqs2, accessible from \url{https://github.com/soedinglab/MMseqs2}.

The custom code for the COMPASS-PTM framework, including the implementation of the two-stage coarse-to-fine process, is available at \textcolor{red}{\url{https://github.com/ZhangJJ26/COMPASS-PTM}}.

\section*{Author contributions}
J.Z. and H.C. conceptualized the study, developed the methodology, implemented the model, performed the analyses, and wrote the original draft. Z.G. and Y.W. contributed to the model's implementation and the experimental design. S.L. and J.X. performed data curation. C.T. and J.Z. conducted the literature investigation and validated the analytical methods. C.H., C.G., and P.H. supervised the project and contributed to the writing, review, and editing of the manuscript. All authors discussed the results and contributed to the final manuscript.

\section*{Competing interests}
The authors declare no competing interests. 

\bibliography{reference.bib}

\begin{appendices}

\section{Experimental Setting}\label{secA1}
\subsection{Datasets}
\label{secA1:dataset}

\begin{figure}[!htbp]
    \centering
\includegraphics[width=1.0\textwidth]{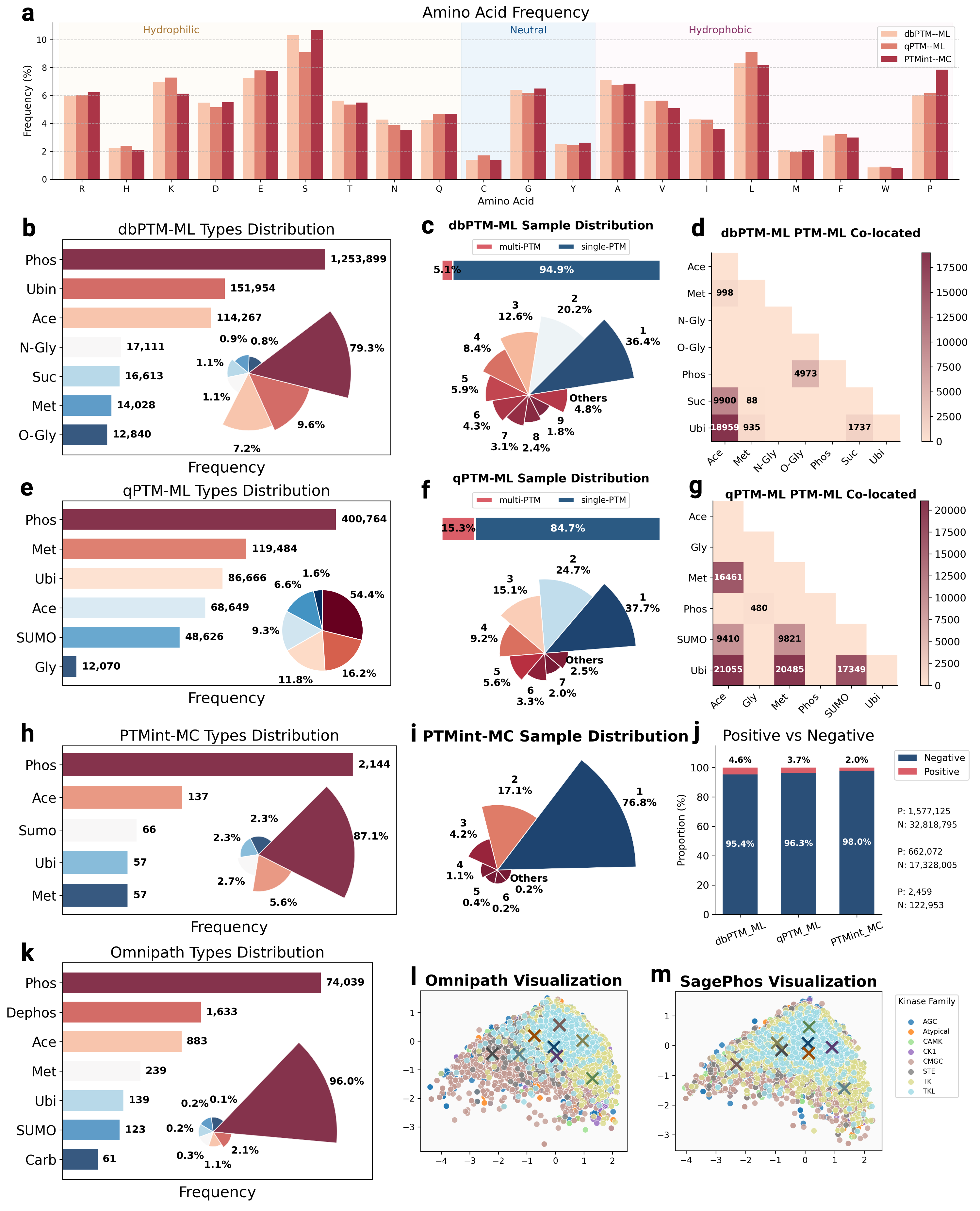}
\caption{\textbf{Overview of the Benchmark Datasets.} \textbf{a}, A comparison of amino acid frequencies in the main datasets. \textbf{b-d}, Key statistics for the dbPTM-ML dataset, showing (\textbf{b}) the frequency of different PTM types, (\textbf{c}) the ratio of single-PTM to multi-PTM sites with a further breakdown of how many sites have 2, 3, or more PTMs, and (\textbf{d}) how often different PTMs are found together at the same site. \textbf{e-g}, The same statistical analysis performed on the qPTM-ML dataset. \textbf{h}, \textbf{i}, PTM type and sample distributions for the single-label PTMint-MC dataset. \textbf{j}, The ratio of modified (Positive) to unmodified (Negative) sites, highlighting the large class imbalance in the three datasets. \textbf{k-m}, A look at the datasets used for stage2 enzyme-substrate prediction, showing (\textbf{k}) PTM type frequencies in Omnipath, and (\textbf{l}, \textbf{m}) a PCA visualization of the initial substrate embeddings, colored by their corresponding kinase family, for the Omnipath (\textbf{l}) and SagePhos (\textbf{m}) datasets.}
    \label{Data}
\end{figure}

In this two-stage process, we employ distinct datasets for different phases and tasks, as shown in Fig. \ref{Data}. 

\subsubsection{priliminaries}
\textbf{dbPTM} database\cite{chung2025dbptm}, a long-standing integrated resource dedicated to Post Translational Modifications (PTMs) research, has been systematically maintained for over a decade to support functional and structural analyses of PTMs. The 2025 update of dbPTM represents a substantial expansion, now cataloging more than 2.79 million PTM sites across diverse proteomes. Of these, approximately 2.243 million sites are experimentally validated, curated from 48 established databases and over 80,000 peer-reviewed research articles through rigorous manual extraction. This extensive compilation ensures the database’s utility as a foundational platform for investigating PTM-mediated regulatory mechanisms.

\textbf{qPTM} database \cite{yu2023qptm} serves as a repository for quantitative PTM proteomics data, addressing the critical need for integrated analysis of reversible PTM dynamics across biological states. This resource expands upon the earlier qPhos framework to encompass six major PTM types — phosphorylation, acetylation, glycosylation, methylation, SUMOylation, and ubiquitylation — across four model organisms. Curated from 2,596 experimental conditions derived from 553 published studies, qPTM integrates 11,482,533 quantitative events mapping to 660,030 non-redundant PTM sites on 40,728 proteins, with matched global proteome data incorporated where available.

\textbf{PTMint} database\cite{hong2023ptmint} serves as a manually curated repository for experimentally validated PTM-mediated regulatory events on protein-protein interactions (PPIs). This resource specifically catalogs PTM sites that either enhance (increase affinity) or inhibit (decrease affinity) PPIs across six model organisms. Its inaugural release integrates 2,477 non-redundant PTM sites mapped to 1,169 proteins, which collectively modulate 2,371 protein-protein pairs implicated in 357 human diseases.

\textbf{PTMCode 2}\cite{minguez2015ptmcode} is an integrative resource documenting functional associations between PTMs within and across protein complexes. It incorporates 316,546 experimentally verified PTM sites spanning 69 modification types across 19 eukaryotic species. Through orthology-based propagation, these annotations expand to $\geq1.6$ million predicted PTM sites, enabling the exploration of 17 million functional PTM-PTM associations involving $\geq100,000$ proteins. This framework provides the most comprehensive landscape of context-specific PTM regulatory networks to date.

\textbf{OmniPath}\cite{turei2016omnipath} is a meta-database that integrates molecular prior knowledge from more than 100 publicly available resources to enable systems-level analysis of cellular regulation. Its core innovation lies in standardizing fragmented biological data into five interoperable modules: signaling networks, annotated protein complexes, multi-attribute protein annotations, intercellular communication roles, and critically — a comprehensive repository of enzyme-substrate relationships governing PTMs. The PTM module, curated from 11 specialized resources, defines 39,201 enzyme-substrate interactions involving 1,821 enzymes and 16,467 substrate-specific modification sites across humans, mice, and rats. 

\textbf{UniprotKB}\cite{boutet2016uniprotkb} The UniProt Knowledgebase (UniProtKB) serves as a globally recognized resource for comprehensive protein annotation, integrating experimental and computationally inferred data to deliver accurate, standardized functional insights. Comprising two distinct sections, UniProtKB/Swiss-Prot provides manually curated records enriched with literature-derived evidence and expert-evaluated computational predictions, while UniProtKB/TrEMBL offers computationally analyzed entries awaiting full manual curation. The latest release of UniProtKB/Swiss-Prot encompasses 573,230 rigorously annotated protein entries, each detailing sequence information, taxonomic classification, functional roles, structural features, and disease associations. 

\textbf{SAGEPhos}\cite{zhang2025sagephos} dataset integrates phosphorylation sites from major repositories, including Phospho.ELM, PhosphoNetworks, and PhosphoSitePlus, augmented by structural data derived from AlphaFoldDB to enhance prediction accuracy. Following stringent quality control and redundancy removal, the dataset encompasses 18,360 high-confidence positive phosphorylation sites, representing a robust foundation for training and validating computational models. 

\subsubsection{MSPN datasets}
\label{secA1.2:MSPN}
In the first stage, we introduced three datasets for PTM site prediction: two novel multi-label datasets curated from dbPTM and qPTM, termed dbPTM-ML and qPTM-ML respectively, and one multi-class dataset derived from PTMint, designated as PTMint-MC.

\textbf{dbPTM-ML} To address the dual long-tail challenges inherent in multi-label PTM prediction—where both PTM type distribution and site-specific positive/negative sample ratios exhibit extreme skewness (Fig. \ref{Data} j) —we implemented stringent data curation criteria. Only PTM types exceeding tens of thousands of experimentally documented sites were retained, ensuring statistically robust representation across both dominant and rare modification categories. This curation yielded seven high-confidence PTM categories with substantial experimental coverage: phosphorylation, acetylation, ubiquitylation, N-linked glycosylation, O-linked glycosylation, methylation, and succinylation, while collectively classifying all other modification types into a consolidated rare category to address long-tail distribution challenges (Fig. \ref{Data}b). This process resulted in a dataset where a fraction of sites contains multiple PTMs, and the specific co-occurrence patterns were characterized (Fig. \ref{Data}c, d). 
As the dbPTM database lacks full-length protein sequences, we retrieved canonical sequences from UniProt \cite{uniprot2023uniprot} to contextualize PTM sites within native structural environments. To preserve local sequence contexts critical for PTM recognition while optimizing computational efficiency, we developed a greedy segmentation algorithm. Although conceptually analogous to sliding window approaches, our method fundamentally differs through its adaptive fragment selection mechanism: rather than fixed-step window sliding, the algorithm dynamically identifies optimal cleavage boundaries to generate peptides $\leq50$ aa that mandatorily contain $\geq1$ PTM site. This constraint guarantees functional relevance while eliminating non-informative segments, yielding a final curated repository of 1,577,125 PTM sites.
We further mitigated overfitting risks using MMseqs2 \cite{steinegger2017mmseqs2} for sequence-space partitioning. Clustering at 40\% sequence identity threshold generated non-redundant, phylogenetically independent subsets: 80\% training, 10\% validation, and 10\% testing. This strategy minimizes bias from evolutionary correlations during model evaluation.

\textbf{qPTM-ML} dataset integrates all six PTM types documented in the qPTM database—phosphorylation, acetylation, ubiquitylation, glycosylation, methylation, and SUMOylation—with each modification type exhibiting robust experimental coverage (Fig. \ref{Data}e). Consequently, no rare-class designation was required. Identical data processing pipelines were implemented for peptide segmentation and sequence partitioning as described for dbPTM-ML, yielding a final curated repository of 662,072 PTM sites, whose multi-label and co-occurrence distributions are detailed in Fig. \ref{Data}f and Fig. \ref{Data}g, respectively.

\textbf{PTMint-MC} Given the limited scale of PTMint (2,477 non-redundant sites) and the near-absence of co-occurring PTM labels at identical sites, we exclusively allocated the entire dataset to a unified test cohort for standalone evaluation  (Fig. \ref{Data}h, i). Processed through the same peptide segmentation pipeline as dbPTM-ML, these data were structured as a multi-class classification task, serving as an independent benchmark for model inference of context-specific PTM functional impacts.

\subsubsection{ESPS datasets}
\label{secA1.3:ESPS}
For the second stage, we focused on the Enzyme-Substrate Pairing (ESP) task. To this end, we constructed and utilized two distinct datasets to train and evaluate the model's ability to predict specific enzyme-substrate interactions. Both datasets are structured with peptides centered on the modification site.

Our primary dataset was constructed using enzyme-substrate relationships curated from the OmniPath database \cite{turei2016omnipath}, with full-length protein sequences for both enzymes and substrates retrieved from UniProt. For this dataset, substrate sequences were processed into 15-mer peptides centered on the modification site. To rigorously assess the model's performance and generalization capabilities, we designed three distinct evaluation settings: a warm-start scenario and two cold-start scenarios. In the warm-start setting, both enzymes and substrate peptides could appear in both the training and test sets, evaluating the model's performance on familiar data. To test generalization to novel entities, we implemented two stringent cold-start splits: a substrate cold-start setting, where all substrate peptides in the test set were strictly excluded from the training set, and an enzyme cold-start setting, where all enzymes present in the test set were entirely absent from the training set. For all settings, negative samples were generated by selecting unmodified residues of the same amino acid type from within the same substrate peptide, maintaining a balanced 1:1 positive-to-negative sample ratio.

For comprehensive validation, we also utilized the established SAGEPhos benchmark dataset \cite{zhang2025sagephos}. This dataset provides full-length kinase sequences paired with 11-mer substrate peptides, where the central residue represents the potential phosphorylation site. 

The use of these two complementary datasets, with their differing sequence lengths and curation strategies, allows for a thorough and robust assessment of our model's ability to learn the molecular recognition patterns that govern enzyme-substrate specificity.

\subsection{Evaluation Metric}
\label{secA3:metric}

To comprehensively evaluate the performance of our proposed model, we employed a diverse set of metrics that capture different aspects of classification quality and predictive capability.

\textbf{Accuracy} measures the proportion of correctly predicted instances (both true positives and true negatives) among all predictions. It is calculated as:

\begin{equation}
Accuracy = \frac{TP + TN}{TP + TN + FP + FN}
\end{equation}

where TP (True Positives) represents correctly identified positive instances, TN (True Negatives) represents correctly identified negative instances, FP (False Positives) represents negative instances incorrectly classified as positive, and FN (False Negatives) represents positive instances incorrectly classified as negative. While straightforward, accuracy alone may not be informative in imbalanced datasets.

\textbf{Precision} (positive predictive value) quantifies the proportion of true positive predictions among all positive predictions, reflecting the model's ability to avoid false positives. The formula for precision is:

\begin{equation}
Precision = \frac{TP}{TP + FP}
\end{equation}

which measures how many of the samples predicted as positive are actually positive.

\textbf{Recall} (sensitivity) measures the proportion of actual positive instances correctly identified by the model, indicating its ability to detect PTM sites without missing true positives. It is calculated as:

\begin{equation}
Recall = \frac{TP}{TP + FN}
\end{equation}

quantifying how completely the model captures all positive instances in the dataset.

\textbf{F1-score} represents the harmonic mean of precision and recall, providing a balanced measure that is particularly useful for evaluating performance on imbalanced datasets where positive PTM sites are significantly fewer than negative sites. The F1-score is defined as:

\begin{equation}
F1 = 2 \times \frac{Precision \times Recall}{Precision + Recall}
\end{equation}

effectively balancing the trade-off between precision and recall.

\textbf{Matthews Correlation Coefficient (MCC)} serves as a balanced measure that considers all four confusion matrix categories. It is calculated as:

\begin{equation}
MCC = \frac{TP \times TN - FP \times FN}{\sqrt{(TP + FP)(TP + FN)(TN + FP)(TN + FN)}}
\end{equation}

producing a value between -1 and +1, where +1 indicates perfect prediction, 0 indicates random prediction, and -1 indicates complete disagreement between prediction and observation. MCC is especially valuable for evaluating binary classifiers on imbalanced datasets as it remains informative even when class distributions vary significantly.

\textbf{Area Under the Receiver Operating Characteristic curve (AUC-ROC)} evaluates the model's ability to discriminate between positive and negative classes across various classification thresholds. The ROC curve plots the true positive rate (TPR) against the false positive rate (FPR) at different threshold settings:

\begin{equation}
TPR = \frac{TP}{TP + FN}
\end{equation}

\begin{equation}
FPR = \frac{FP}{FP + TN}
\end{equation}

AUC-ROC values range from 0 to 1, with higher values indicating better discrimination capability. This metric is particularly valuable as it is insensitive to class imbalance and provides a comprehensive assessment of the model's classification performance across all possible decision thresholds.

\textbf{Area Under the Precision-Recall curve (AUC-PRC)} evaluates the trade-off between precision and recall across all thresholds. The PR curve plots precision against recall (which is equivalent to TPR). Like AUC-ROC, its value ranges from 0 to 1, with higher scores indicating better performance.

These metrics collectively provide a comprehensive evaluation framework, allowing us to assess both the overall prediction accuracy and the model's performance in correctly identifying the typically underrepresented positive PTM sites.

\subsection{Network Architecture}
\label{appendix:architecture}

\begin{figure}[htp]
    \centering
\includegraphics[width=1.0\textwidth]{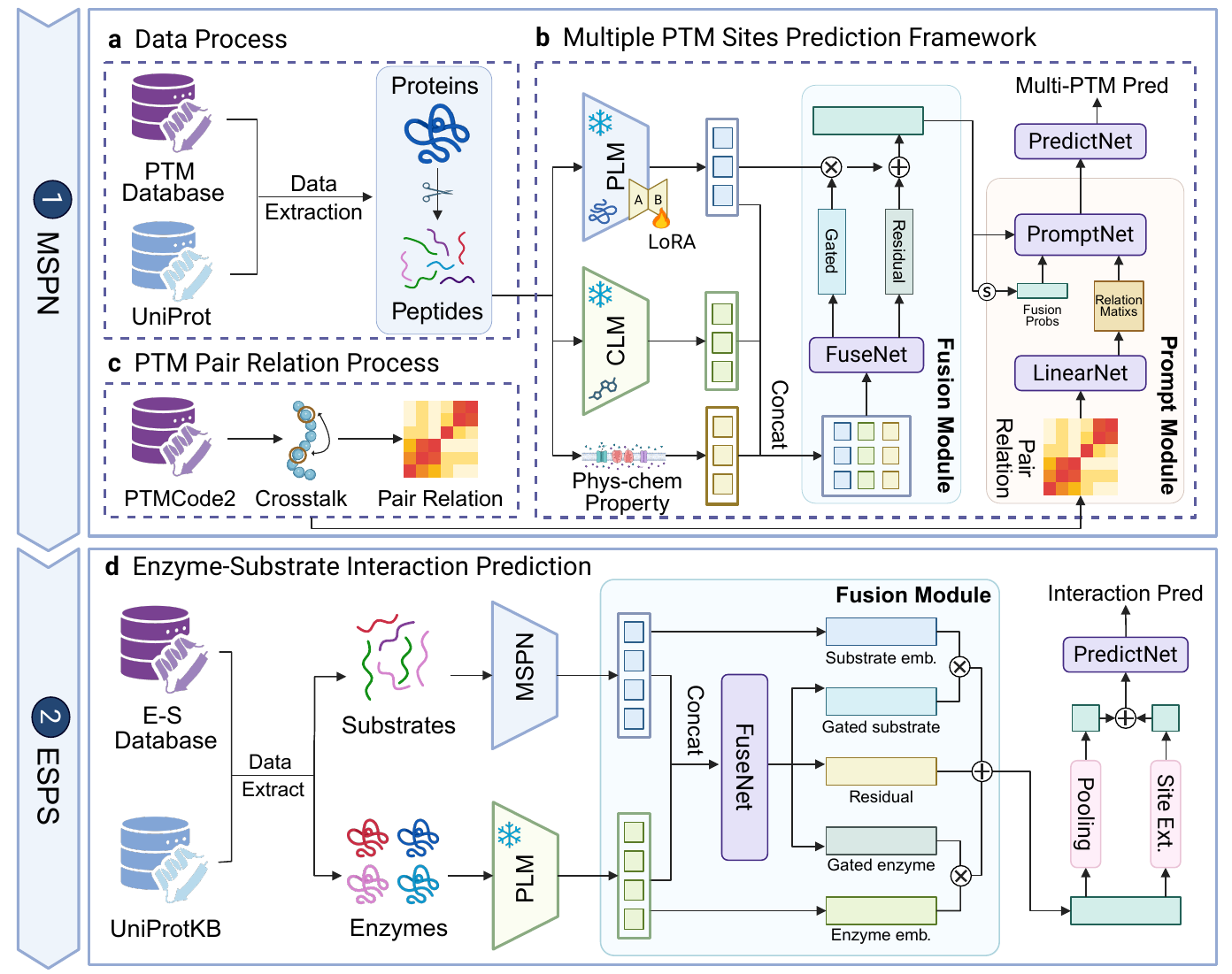}
\caption{\textbf{Details of COMPASS-PTM.} \textbf{a-c}, Components of the first-stage Multi-label Site Profiling Network (MSPN). \textbf{a}, Data processing workflow showing extraction of PTM site annotations from curated PTM databases and full-length protein sequences from UniProt, followed by sequence segmentation into peptides. \textbf{b}, Multi-label PTM site prediction framework comprising protein language model (PLM) and chemical language model (CLM) encoders, LoRA-based fine-tuning, fusion modules for multimodal integration, and prompt-guided prediction networks that output site-specific PTM probabilities. \textbf{c}, PTM crosstalk relationship extraction from PTMCode2 database, processed into learnable pairwise relationship matrix that informs the prompt module for enhanced prediction accuracy. \textbf{d}, Second-stage Enzyme-Substrate Pairing System (ESPS) utilizing substrate and enzyme embeddings and fusion module to predict catalytic compatibility.}
    \label{COMPASS-PTM_appendix}
\end{figure}

Our COMPASS-PTM model is built upon a dual-modal architecture that integrates evolutionary, structural, and chemical information to predict Post-Translational Modification (PTM) sites, as shown in Fig. \ref{COMPASS-PTM_appendix}

\subsubsection{Dual-Modal Encoders and Feature Representation}
The model foundation consists of two complementary language models. For biological context, we use a Protein Language Model (PLM) based on ESM-2 (esm2\_t30\_150M\_UR50D) \cite{lin2023evolutionary}, which features 30 transformer layers, 20 attention heads per layer, and produces 640-dimensional embeddings. To capture the chemical properties of amino acids, we employ a Chemical Language Model (CLM) based on SELF-BART model \cite{priyadarsini2024selfbart}. In addition to these learned representations, we incorporate explicit physicochemical features, which provides a 4-dimensional vector for each amino acid encoding its molecular weight, isoelectric point, hydrophobicity, and polarity. To efficiently adapt the powerful PLM to the PTM prediction task, we employ Low-Rank Adaptation (LoRA) \cite{hu2022lora}. We apply LoRA with a rank of $r=16$ and a scaling factor of $\mathbf{\alpha}=16$ to the query, key, and value matrices of the PLM's attention layers. A dropout rate of 0.1 is used within the LoRA modules to prevent overfitting. This strategy allows for efficient fine-tuning while keeping the base PLM's parameters frozen, preventing catastrophic forgetting (Fig. \ref{COMPASS-PTM_appendix}b).

\subsubsection{Crosstalk-Aware Prompting and Feature Fusion}
The outputs from the PLM, CLM, and physicochemical features are first combined by a fusion network into a unified 640-dimensional representation. This fused embedding is then processed by two specialized transformer layers designed to model PTM interdependencies via crosstalk-aware prompting. Our custom attention mechanism within these layers first generates preliminary PTM probabilities via a softmax-activated head. Concurrently, a crosstalk prior matrix (Fig.\ref{COMPASS-PTM_appendix}c) is transformed through two learnable linear projections. An attention bias is then computed by modeling the interaction between the PTM probabilities and the transformed crosstalk matrix. This bias is added to the standard scaled dot-product attention scores, effectively steering the model's attention towards biochemically plausible PTM relationships.

\subsubsection{Classification Head and Training Details}
The final representation is passed to a Residual MLP classification head, which outputs logits for the 8 PTM types. The model was trained using the Adam optimizer \cite{kingma2014adam} (learning rate \(2\times10^{-5}\)). We used a batch size of 256, and trained for a maximum of 100 epochs. These hyperparameters were uniformly applied across all model variations, including different ESM2 backbones (8M, 35M, 150M, and 650M). All experiments were conducted on a single NVIDIA A800 GPU.

\section{Supplementary Results}\label{secB1}

\subsection{Detailed Cross-Task Benchmark Against PTMGPT2}
As stated in the main text, to rigorously evaluate the generalizability of the principles embodied in our architecture, we conducted a challenging cross-task benchmark against PTMGPT2, a strong single-PTM predictor. For the selected five binary classification datasets used in the PTMGPT2 study, we retrained MSPN from scratch using only the same dataset, without carrying over weights or performing any task-specific fine-tuning. This deliberately disadvantageous setting was chosen to probe the robustness of the model’s learned representations under varied data distributions.

Despite the differences in training objectives, the results unequivocally demonstrate the broad generalizability of COMPASS-PTM (Fig. 2\textbf{d}). Across all five datasets, the model achieved substantially more balanced and accurate performance profiles. For example, on Lysine Acetylation, COMPASS-PTM achieved an F1-Score of 0.800 and an MCC of 0.461, exceeding the baseline’s scores of 0.406 and 0.221 by 96.8\% and 108.6\%, respectively. The robustness of our approach is further underscored by its outstanding performance on Lysine Hydroxylation—a relatively rare modification with distinct biochemical properties—where MSPN achieved an F1-Score of 0.885, representing a 29.8\% improvement over PTMGPT2.

COMPASS-PTM's high performance on foundational tasks demonstrates its capacity to learn the intrinsic rules of PTMs rather than spurious, task-specific correlations. This inherent robustness validates our architecture and establishes COMPASS-PTM as a powerful and versatile predictive engine.

\begin{figure}[!htp]
    \centering
\includegraphics[width=1.0 \textwidth]{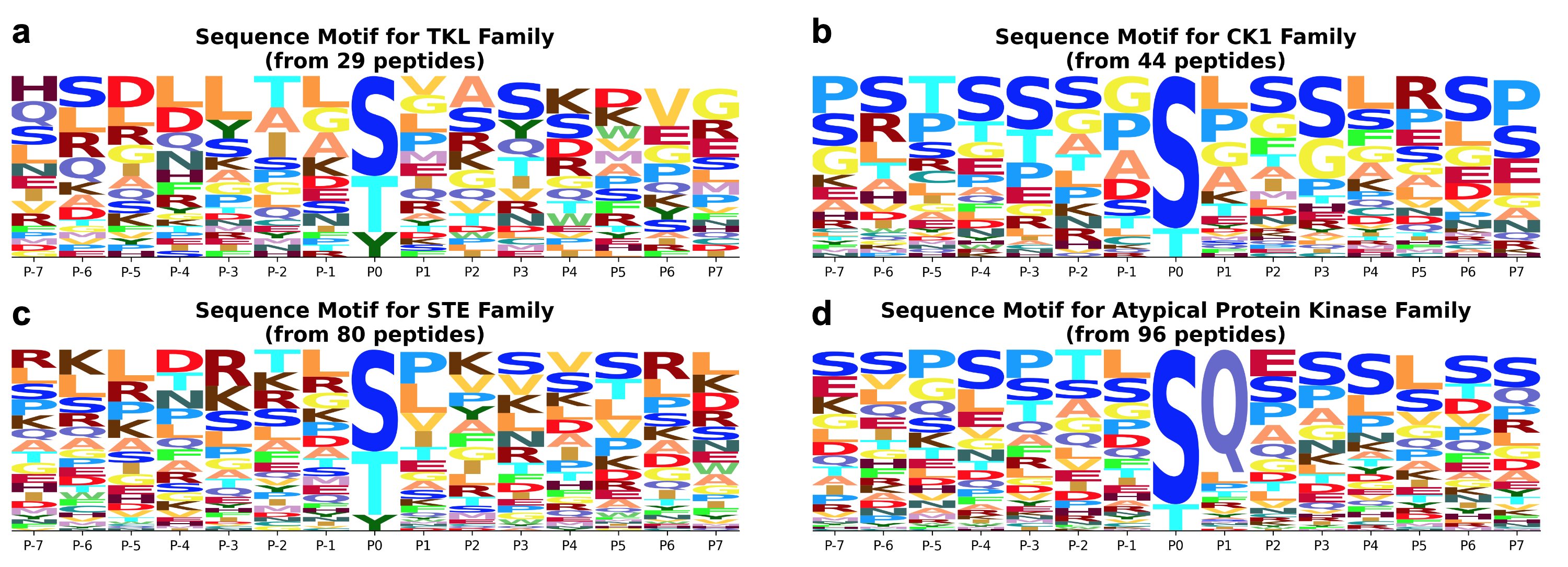}
\caption{\textbf{Substrate Motif Recognition for Additional Kinase Families.} \textbf{a-d}, Sequence logos generated from high-confidence substrate peptides predicted by COMPASS-PTM for four additional kinase families: TKL (\textbf{a}, n=29 peptides), CK1 (\textbf{b}, n=44), STE (\textbf{c}, n=80), and Atypical protein kinases (\textbf{d}, n=96). }
    \label{sup_results}
\end{figure}

\subsection{Detailed Zero-Shot Study to Unseen Kinases}
\subsubsection{Performance Comparison on the DARKIN Benchmark}
To rigorously assess the zero-shot generalization capability of our model, we evaluated it on the DARKIN benchmark\cite{sunar2025darkin}. The DARKIN benchmark is a zero-shot learning framework designed to evaluate a model's ability to associate phosphosites with previously unseen ``dark kinases"—kinases that are understudied. This task is of significant biological importance, as experimentally identifying the specific kinase for over 95\% of known phosphosites remains a major challenge , and kinases are crucial drug targets in many diseases

The benchmark's integrity is ensured through a carefully designed data splitting strategy. It enforces strictly disjoint sets of kinases between its training, validation, and testing splits. Furthermore, to prevent overly optimistic results and ensure a true zero-shot evaluation, the splits are stratified by kinase groups and kinases with high sequence similarity are exclusively assigned to the same set. Performance is measured using macro Average Precision (mAP), a metric chosen for its stability and robustness against class imbalance, which is common in this problem.

We compare our model, COMPASS-PTM, against the top-performing protein language model encoders and fine-tuning strategies reported in the original DARKIN study. As shown in Table~\ref{tab:darkin_comparison}, COMPASS-PTM substantially outperforms all prior methods. The previous state-of-the-art performance on this challenging benchmark was an mAP of 0.1911, achieved by a fine-tuned ESM1B model that was augmented with explicit biological features (kinase family, group, and EC information). In stark contrast, our model, COMPASS-PTM, achieves a new state-of-the-art mAP of 0.2946, surpassing the best baseline by a remarkable 54\%.

Notably, our model achieves this superior performance while using a significantly more parameter-efficient ESM2-150M backbone. This result is particularly significant given that the DARKIN study itself concludes that dark kinase prediction remains a ``highly challenging task" for current protein language models. The substantial performance gain underscores the effectiveness of COMPASS-PTM's architecture, which learns representations highly informative for kinase-substrate specificity through an initial stage dedicated to learning PTM-specific codes.

\begin{table}[htp!]
\captionsetup{width=\textwidth}
\centering
\caption{Zero-shot performance on the DARKIN benchmark, evaluated using mean Average Precision (mAP). The best-performing model is highlighted in \textbf{bold}; the previous state-of-the-art is \textit{italicized}.}
\label{tab:darkin_comparison}
\begin{tabularx}{0.75\textwidth}{>{\raggedright\arraybackslash}X c S[table-format=1.4]}
\toprule
\textbf{Model} & \textbf{Parameters} & {\textbf{mAP}} \\
\midrule
\multicolumn{3}{l}{\textit{Baseline Sequence Encoders}} \\
ESM2 (avg) & 650M & 0.1391 \\
ProtT5-XL & 3B & 0.1552 \\
ESM1B (cls) & 650M & 0.1631 \\
ESM1v (cls) & 650M & 0.1640 \\
\midrule
\multicolumn{3}{l}{\textit{Encoder with Additional Kinase Features}} \\
SaProt (cls) + Features & 650M & 0.1800 \\
\midrule
\multicolumn{3}{l}{\textit{Encoder with Fine-tuning}} \\
ProtT5-XL + Fine-tuning + Features & 3B & 0.1800 \\
ESM1B (cls) + Fine-tuning + Features & 650M & \textit{0.1911} \\
\midrule
\textbf{COMPASS-PTM (Ours)} & \textbf{150M} & \textbf{0.2946} \\
\bottomrule
\end{tabularx}
\end{table}

\subsubsection{Evaluation Metrics}
The model's performance on the zero-shot kinase-substrate prediction task was quantitatively assessed using the mean Average Precision (mAP), which provides a comprehensive measure of the model's ability to rank the true positive kinases higher than the negative ones across all substrates in the test set.

For a dataset comprising $N$ substrates, the mAP is formulated as the mean of the Average Precision (AP) scores computed for each individual substrate:
\begin{equation}
\label{eq:map}
\text{mAP} = \frac{1}{N} \sum_{i=1}^{N} \text{AP}_i
\end{equation}
where $\text{AP}_i$ is the Average Precision for the $i$-th substrate.

The AP for a single substrate, $s$, is calculated based on the ranked list of all $K$ candidate kinases, ordered by the model's predicted interaction scores. It is defined as the sum of the precision values at each rank corresponding to a true positive kinase, averaged over the total number of true positives for that substrate. The formula is given by:
\begin{equation}
\label{eq:ap}
\text{AP}_s = \frac{1}{|\mathcal{K}_s^+|} \sum_{k=1}^{K} P(k) \cdot \mathbb{I}(\text{rank}_k(s) \in \mathcal{K}_s^+)
\end{equation}
Here, $\mathcal{K}_s^+$ is the set of true positive kinases for substrate $s$, and $|\mathcal{K}_s^+|$ is its cardinality. The term $\text{rank}_k(s)$ denotes the kinase at rank $k$ in the predicted list. The indicator function, $\mathbb{I}(\cdot)$, is 1 if the kinase at rank $k$ is a true positive, and 0 otherwise.

The precision at a given rank $k$, $P(k)$, measures the fraction of true positives among the top-$k$ predictions:
\begin{equation}
\label{eq:precision_at_k}
P(k) = \frac{1}{k} \sum_{j=1}^{k} \mathbb{I}(\text{rank}_j(s) \in \mathcal{K}_s^+)
\end{equation}
This formulation ensures that the AP score rewards models that not only retrieve correct kinases but also place them at higher ranks in the prediction list, providing a robust evaluation of the model's generalization capability in the zero-shot context.

\subsection{Substrate Motif Recognition for Additional Kinase Families}
Our analysis of the remaining kinase families revealed that the model successfully recovered a diverse range of canonical motifs, each with distinct biochemical features. For the TKL (Tyrosine Kinase-Like) Family, a diverse group that structurally resembles tyrosine kinases but phosphorylates serine/threonine, the resulting logo reflects the known diversity of this group; while lacking a single, highly constrained consensus, the model successfully captured several subtle but significant preferences, including an enrichment for basic residues at P-2 and small, non-polar residues at P+1, showing it is capable of learning weaker sequence patterns from functionally diverse enzyme families (Fig. \ref{sup_results}a). The model's capacity to learn complex signatures was further evidenced by the logo for the CK1 (Casein Kinase 1) Family, a highly conserved group of acidophilic kinases involved in numerous cellular processes, which perfectly captures its hallmark feature: a strong preference for a Serine (S) or Threonine (T) at the P-3 position. This reflects the well-established primed acidophilic motif (\emph{S/T-X-X-S/T}) for CK1, where an upstream phosphorylation event is often required for substrate recognition (Fig. \ref{sup_results}b). For the STE (Sterile Homolog) family, which constitutes a major part of the mitogen-activated protein kinase (MAPK) signaling cascades, the model captured a multi-faceted recognition motif. The most dominant feature identified is the stringent requirement for a Proline (P) at the P+1 position, which defines the canonical proline-directed signature (\emph{S/T-P}) essential for MAPK signaling. Additionally, it captured a significant secondary signal for basic amino acids, especially Lysine (K), at the P-2 position, showcasing its ability to learn complex sequence patterns (Fig. \ref{sup_results}c). Finally, for the Atypical Protein Kinase Family, a functionally diverse group that lacks sequence homology to the main protein kinase superfamily, the model accurately learned the \emph{L-S/T-Q} consensus motif, which is the signature of the clinically significant PIKK subgroup (including mTOR, ATM, and ATR). This highlights the model's ability to capture the critical preference for a Leucine (L) at P-1 and a Glutamine (Q) at P+1 that governs recognition by these kinases (Fig. \ref{sup_results}d). 

\subsection{More Cases for Predicting PTM Alterations Driven by Pathogenic
Mutations}
\label{secB3}

In addition to the cases presented in the main text, we further highlight COMPASS-PTM's utility by examining several cancer-associated somatic mutations whose functional impacts have recently been computationally validated by the DeepMVP study \cite{wen2025deepmvp}.

\begin{figure}[!htp]
    \centering
\includegraphics[width=1.0 \textwidth]{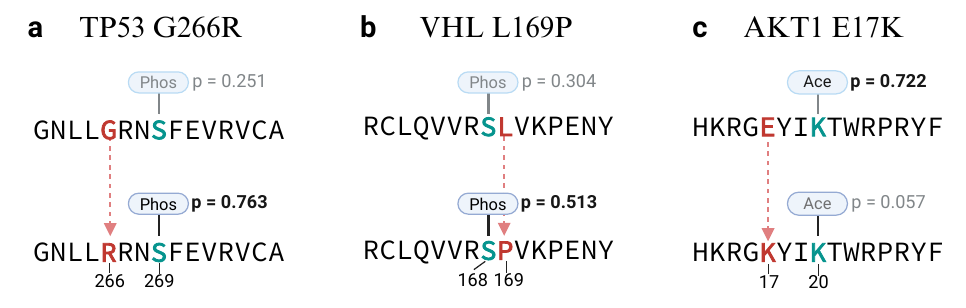}
\caption{\textbf{More case of predicting the PTM consequences of pathogenic variants.} a, The TP53 p.G266R cancer-associated substitution is predicted to induce a gain of phosphorylation at the proximal S269 site (predicted probability increases from 0.251 to 0.763). b, The VHL p.L169P variant, implicated in renal carcinoma, is predicted to cause a gain of phosphorylation at the adjacent S168 site (probability increases from 0.304 to 0.513). c, The AKT1 p.E17K oncogenic variant is predicted to result in a substantial loss of acetylation at the nearby K20 site (probability drops from 0.722 to 0.057). For each case, the diagrams illustrate the local sequence and predicted PTM probabilities, where the pathogenic substitution is highlighted in red and the affected PTM site is in green.}
    \label{sup:mutation}
\end{figure}

For the TP53 p.G266R loss-of-function variant, COMPASS-PTM generated the novel hypothesis that the mutation induces a gain of phosphorylation at S269 (Fig. \ref{sup:mutation}a). This offers a compelling explanation for the variant's pathogenic effect, as S269 phosphorylation is reported to inhibit TP53 activity.
Similarly, for the AKT1 p.E17K hotspot substitution, a known driver of oncogenic activation, our model predicted a substantial loss of acetylation at the adjacent K20 site (Fig. \ref{sup:mutation}c). This prediction aligns perfectly with experimental evidence that K20 acetylation inhibits AKT1 activity, thus providing a direct mechanistic link where the E17K variant activates the kinase by reducing this inhibitory modification.
Furthermore, the model proposed a functional consequence for the recurrent but uncharacterized VHL p.L169P variant, predicting a gain of phosphorylation at the S168 site (Fig. \ref{sup:mutation}b). This suggests the variant may contribute to tumorigenesis by promoting an S168 phosphorylation event that is known to mark the VHL tumor suppressor for ubiquitination and degradation. 

These examples further underscore the utility of COMPASS-PTM in generating precise, testable hypotheses to elucidate the molecular mechanisms underlying pathogenic mutations.

\subsection{Predicted PTM-Abolishing Variants in Sperm Proteins}
To demonstrate the utility of COMPASS-PTM in elucidating disease mechanisms, we first performed a systematic screen to identify pathogenic mutations with a high likelihood of altering the post-translational code. To probe the genetic basis of reproductive pathology, a critical yet understudied field, we cross-referenced pathogenic variants from the PhosphoSitePlus PTMVar database \cite{hornbeck2015phosphositeplus} against a comprehensive catalog of the human sperm proteome \cite{castillo2018contribution}. By applying a stringent filtering criterion (wild-type prediction score \textgreater{} 0.8 and mutant prediction score \textless{} 0.2) to pinpoint mutations predicted to abolish PTM events, our screen identified 62 high-confidence SNPs (Table \ref{tab:ptm_variants}).

\begin{footnotesize}
\setlength{\tabcolsep}{3pt}

\begin{longtable}{@{} l l c r c c l S[table-format=1.3] S[table-format=1.3] S[table-format=-1.3] @{}}
\caption{Predicted impact of 62 genetic variants on PTM sites. WT\_Prob. and MT\_Prob. represent the predicted probabilities for the wild-type and mutant sequences, respectively. Diff is the difference score(MT\_Probab - WT\_Probab). Abbreviations: Phos, Phosphorylation; Ubiq, Ubiquitylation; Meth, Methylation; SUMO, Sumoylation.} \label{tab:ptm_variants}\\

\toprule
\textbf{Gene} & \textbf{\begin{tabular}{@{}c@{}}UniProt\\ ID\end{tabular}} & \textbf{WT} & \textbf{Pos.} & \textbf{Var} & \textbf{\begin{tabular}{@{}c@{}}Mod.\\ Pos.\end{tabular}} & \textbf{PTM Type} & {\textbf{\begin{tabular}{@{}c@{}}WT\\ Prob.\end{tabular}}} & {\textbf{\begin{tabular}{@{}c@{}}MT\\ Prob.\end{tabular}}} & {\textbf{Diff}} \\
\midrule
\endfirsthead

\caption{ (Continued) Predicted impact of 62 genetic variants on PTM sites.}\\
\toprule
\textbf{Gene} & \textbf{\begin{tabular}{@{}c@{}}UniProt\\ ID\end{tabular}} & \textbf{WT} & \textbf{Pos.} & \textbf{Var} & \textbf{\begin{tabular}{@{}c@{}}Mod.\\ Pos.\end{tabular}} & \textbf{PTM Type} & {\textbf{\begin{tabular}{@{}c@{}}WT\\ Prob.\end{tabular}}} & {\textbf{\begin{tabular}{@{}c@{}}MT\\ Prob.\end{tabular}}} & {\textbf{Diff}} \\
\midrule
\endhead

\bottomrule
\multicolumn{10}{r}{\textit{Continued on next page}}
\endfoot

\bottomrule
\endlastfoot

APC & P25054 & S & 2621 & C & 2621 & Phos & 0.944 & 0.000 & -0.944 \\
HUWE1 & Q7Z6Z7 & K & 4295 & N & 4295 & Ubiq & 0.869 & 0.000 & -0.869 \\
HEXA & P06865 & K & 342 & R & 342 & Ubiq & 0.820 & 0.000 & -0.820 \\
MAP2K1 & Q02750 & K & 57 & E & 57 & Ubiq & 0.843 & 0.000 & -0.843 \\
USP8 & P40818 & S & 718 & P & 718 & Phos & 0.931 & 0.000 & -0.931 \\
GIGYF2 & Q6Y7W6 & S & 273 & C & 273 & Phos & 0.865 & 0.000 & -0.865 \\
MITF & O75030 & E & 425 & K & 423 & SUMO & 0.863 & 0.000 & -0.862 \\
SCN1A & P35498 & S & 525 & F & 525 & Phos & 0.869 & 0.000 & -0.869 \\
ACP1 & P24666 & S & 137 & F & 137 & Phos & 0.867 & 0.000 & -0.867 \\
ACE2 & Q9BYF1 & K & 26 & R & 26 & Ubiq & 0.855 & 0.000 & -0.855 \\
TNFRSF10A & O00220 & K & 443 & N & 443 & Ubiq & 0.815 & 0.000 & -0.814 \\
FLNC & Q14315 & S & 1624 & L & 1624 & Phos & 0.889 & 0.000 & -0.889 \\
ATP10D & Q9P241 & T & 43 & I & 43 & Phos & 0.880 & 0.000 & -0.880 \\
LRRFIP1 & Q32MZ4 & S & 68 & C & 68 & Phos & 0.917 & 0.000 & -0.917 \\
ZNF592 & Q92610 & T & 1024 & N & 1024 & Phos & 0.904 & 0.000 & -0.904 \\
HIP1R & O75146 & K & 404 & Q & 404 & Ubiq & 0.821 & 0.000 & -0.821 \\
WRN & Q14191 & S & 1133 & A & 1133 & Phos & 0.937 & 0.000 & -0.937 \\
TRIP11 & Q15643 & T & 1846 & I & 1846 & Phos & 0.956 & 0.000 & -0.956 \\
RREB1 & Q92766 & S & 1499 & Y & 1499 & Phos & 0.818 & 0.100 & -0.717 \\
MAP1B & P46821 & S & 1400 & G & 1400 & Phos & 0.889 & 0.000 & -0.889 \\
LATS1 & O95835 & S & 204 & G & 204 & Phos & 0.919 & 0.000 & -0.919 \\
GAS2L1 & Q99501 & S & 490 & G & 490 & Phos & 0.912 & 0.000 & -0.912 \\
PBRM1 & Q86U86 & K & 231 & R & 231 & Ubiq & 0.895 & 0.000 & -0.895 \\
RBM20 & Q5T481 & S & 637 & G & 637 & Phos & 0.952 & 0.001 & -0.951 \\
DHX38 & Q92620 & T & 1217 & A & 1217 & Phos & 0.852 & 0.000 & -0.852 \\
KIF17 & Q9P2E2 & S & 369 & R & 369 & Phos & 0.891 & 0.000 & -0.891 \\
KCNH2 & Q12809 & K & 897 & T & 897 & Phos & 0.000 & 0.867 & 0.867 \\
TRMT2A & Q8IZ69 & S & 602 & R & 602 & Phos & 0.843 & 0.000 & -0.843 \\
CLGN & O14967 & S & 579 & F & 579 & Phos & 0.808 & 0.000 & -0.808 \\
BRAF & P15056 & S & 467 & A & 467 & Phos & 0.906 & 0.000 & -0.906 \\
CDK13 & Q14004 & S & 340 & F & 340 & Phos & 0.865 & 0.000 & -0.865 \\
FBXO4 & Q9UKT5 & S & 12 & L & 12 & Phos & 0.917 & 0.000 & -0.917 \\
PKLR & P30613 & R & 359 & C & 354 & Ubiq & 0.824 & 0.125 & -0.699 \\
MYH9 & P35579 & K & 910 & Q & 910 & Ubiq & 0.900 & 0.000 & -0.900 \\
SLC2A11 & Q9BYW1 & K & 469 & E & 469 & Ubiq & 0.870 & 0.000 & -0.869 \\
BIN1 & O00499 & K & 35 & N & 35 & Ubiq & 0.859 & 0.000 & -0.859 \\
SETX & Q7Z333 & S & 2612 & G & 2612 & Phos & 0.837 & 0.000 & -0.837 \\
MAP2K1 & Q02750 & K & 57 & N & 57 & Ubiq & 0.843 & 0.001 & -0.842 \\
LCP2 & Q13094 & S & 410 & C & 410 & Phos & 0.917 & 0.001 & -0.916 \\
CDK11B & P21127 & S & 414 & L & 414 & Phos & 0.907 & 0.000 & -0.907 \\
USP8 & P40818 & S & 718 & C & 718 & Phos & 0.931 & 0.000 & -0.931 \\
DIABLO & Q9NR28 & S & 126 & L & 126 & Phos & 0.930 & 0.000 & -0.930 \\
ABL1 & P00519 & T & 852 & P & 852 & Phos & 0.851 & 0.000 & -0.851 \\
RREB1 & Q92766 & S & 1140 & F & 1140 & Phos & 0.905 & 0.000 & -0.905 \\
ST13P4 & Q8IZP2 & S & 71 & L & 71 & Phos & 0.802 & 0.000 & -0.802 \\
CANT1 & Q8WVQ1 & S & 303 & R & 303 & Phos & 0.867 & 0.000 & -0.867 \\
CDK13 & Q14004 & T & 494 & A & 494 & Phos & 0.950 & 0.000 & -0.950 \\
LATS1 & O95835 & T & 255 & N & 255 & Phos & 0.856 & 0.000 & -0.856 \\
FMR1 & Q06787 & R & 546 & H & 546 & Meth & 0.980 & 0.000 & -0.980 \\
RAD50 & Q92878 & K & 616 & E & 616 & Ubiq & 0.847 & 0.000 & -0.847 \\
BRCA2 & P51587 & T & 582 & P & 581 & Phos & 0.154 & 0.863 & 0.709 \\
MSH6 & P52701 & S & 65 & L & 65 & Phos & 0.942 & 0.000 & -0.942 \\
SPATA31H1 & Q68DN1 & S & 1665 & P & 1665 & Phos & 0.816 & 0.000 & -0.816 \\
VRK3 & Q8IV63 & S & 59 & F & 59 & Phos & 0.889 & 0.000 & -0.889 \\
RAF1 & P04049 & T & 310 & A & 310 & Phos & 0.856 & 0.000 & -0.856 \\
CERT1 & Q9Y5P4 & S & 132 & L & 132 & Phos & 0.843 & 0.000 & -0.843 \\
PPP4R1 & Q8TF05 & S & 593 & N & 593 & Phos & 0.882 & 0.000 & -0.882 \\
RAF1 & P04049 & S & 259 & F & 259 & Phos & 0.872 & 0.000 & -0.872 \\
RAF1 & P04049 & S & 259 & A & 259 & Phos & 0.872 & 0.000 & -0.872 \\
KIF16B & Q96L93 & K & 772 & T & 772 & Ubiq & 0.863 & 0.000 & -0.863 \\
RALA & P11233 & K & 128 & R & 128 & Ubiq & 0.931 & 0.000 & -0.931 \\
NCL & P19338 & P & 68 & L & 67 & Phos & 0.809 & 0.182 & -0.627 \\

\end{longtable}
\end{footnotesize}

\subsection{Detailed Ablation Studies}
\begin{figure}[!htbp]
    \centering
    \includegraphics[width=1.0\linewidth]{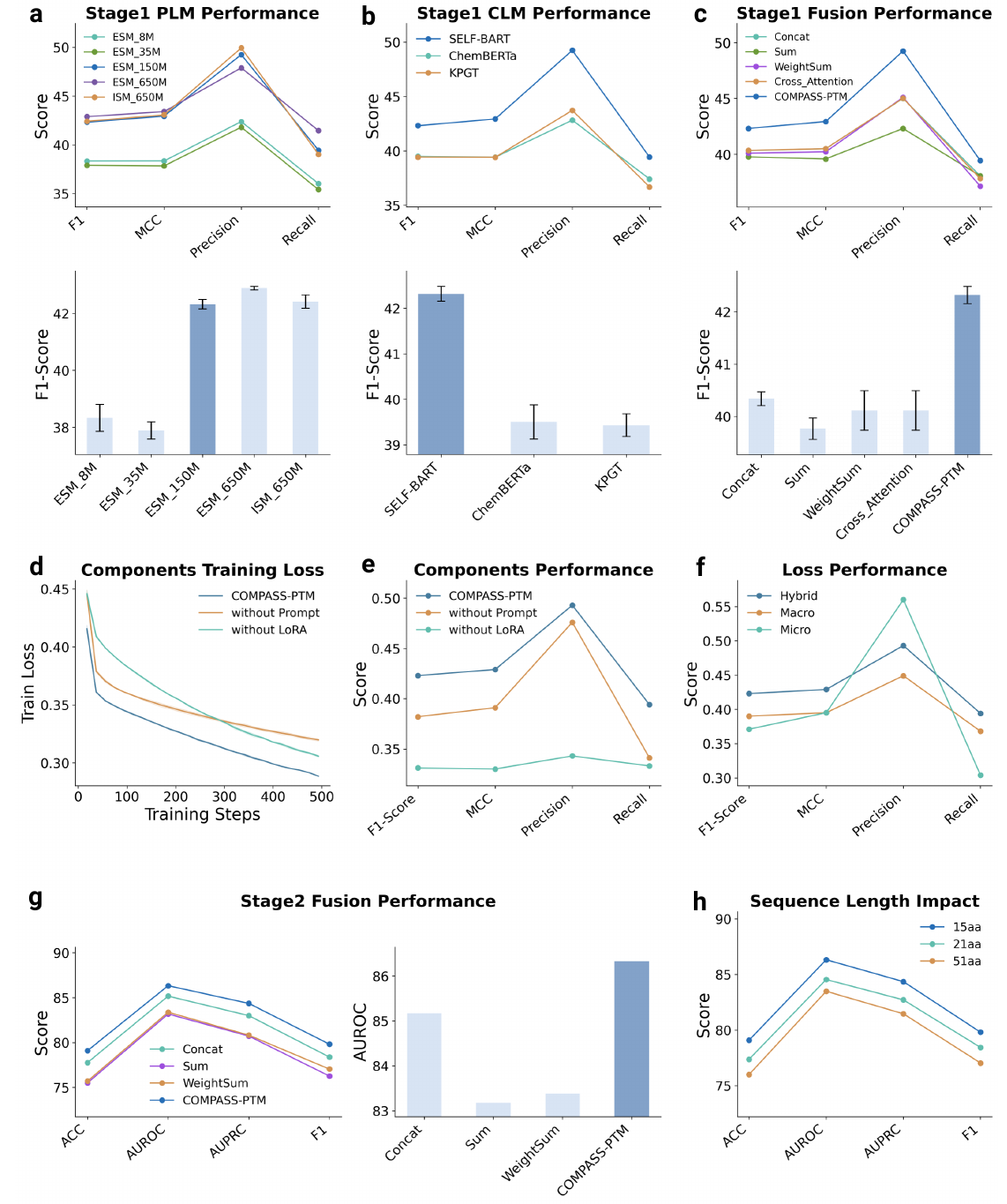}
    \caption{\textbf{Ablation Studies.}\textbf{a-c}, Empirical selection of suitable encoders and the initial fusion strategy at Stage 1. These panels present a comparative analysis to identify the suitable candidates for: (\textbf{a}) the Protein Language Model (PLM) encoder, (\textbf{b}) the Chemical Language Model (CLM) encoder, and (\textbf{c}) the feature fusion architecture. The results from these evaluations guide the component selection for the final model. \textbf{d}, \textbf{e}, Ablation studies investigating the contribution of key architectural components. The panels illustrate (\textbf{d}) the training loss curves and (\textbf{e}) the final performance metrics for the full model versus versions without prompt-tuning and without LoRA, demonstrating their essential role. \textbf{f}, Performance evaluation of the hybrid loss function against standard macro and micro averaging approaches. \textbf{g-h}, Selection of the suitable fusion architecture (\textbf{g}) and input peptide length (\textbf{h}) at Stage 2. For all bar charts, error bars represent the standard deviation calculated over three independent experimental runs.} 
    \label{fig:ablation}
\end{figure}

To systematically validate the design principles of COMPASS-PTM and quantify the contribution of each architectural component, we conducted a series of ablation experiments. These studies evaluate the effectiveness of each designed module, providing a clear rationale for the methodological choices that drive the model's performance.

\subsubsection{Dissecting the Stage 1 Multi-label Site Profiling Network (MSPN)}

\textbf{Optimizing the Dual-Modal Encoder Architecture.} The MSPN's performance is built upon its dual-modal encoder; we therefore began by optimizing its core components. First, for the protein language model, selection experiments revealed that ESM2-150M \cite{lin2023evolutionary} provides the optimal balance between representational power and efficiency. This model improved the F1-score and MCC by 11.6\% and 13.5\% respectively over smaller variants, while larger models offered diminishing returns, as shown in Fig. \ref{fig:ablation}a. Next, evaluation of the chemical language model (Fig. \ref{fig:ablation}b) demonstrated that SELF-BART \cite{priyadarsini2024selfbart} outperformed alternatives (7.1\% F1-score and 8.9\% MCC improvement), validating our hypothesis that physicochemical reactivity patterns provide crucial information complementary to evolutionary context. Finally, our proposed dual-modal fusion architecture proved substantially more effective than conventional methods, outperforming simple concatenation and standard attention by 4.7\% in F1-score, as shown in Fig. \ref{fig:ablation}c. Its gated-residual design effectively treats evolutionary features as primary signals while selectively incorporating chemical information, a principle that enhances discriminative capacity.

\textbf{Contribution of Crosstalk-Aware Prompting.} The crosstalk-aware prompting module is a critical innovation for modeling PTM crosstalk. Ablation experiments (Fig. \ref{fig:ablation}d-e) quantify its substantial impact: its removal results in a 10.7\% decrease in F1-score and a 9.8\% decrease in MCC, accompanied by a degradation in both precision and recall. This confirms that the prompt module successfully resolves ambiguity among competing PTM labels by refining its learned PTM co-occurrence patterns, leading to predictions that better reflect biological reality.

\textbf{Contribution of LoRA Fine-Tuning.} To efficiently adapt the large-scale protein language model to the PTM prediction task, we employed Low-Rank Adaptation (LoRA) \cite{hu2022lora} for parameter-efficient fine-tuning. A direct comparison between models trained with and without LoRA reveals the profound impact of this strategy. Shown in Fig. \ref{fig:ablation}d-e, the use of LoRA boosted key performance metrics by approximately 30\%, including a 44\% improvement in precision. This underscores the necessity of parameter-efficient fine-tuning for effectively specializing large, generalist PLMs to a nuanced task like PTM profiling.

\textbf{Efficacy of the Hybrid Loss Function.} Our hybrid loss function was designed to address the challenging double long-tail distribution of PTM datasets. Ablation experiments confirmed that individual loss functions offer a suboptimal trade-off: Dice loss achieves high recall but low precision, while Focal loss delivers high precision but poor recall. By combining them, our hybrid approach successfully leverages their complementary strengths to achieve a balanced performance profile, culminating in a state-of-the-art F1-score (0.423) and MCC (0.429), as shown in Fig. \ref{fig:ablation}f. This ensures both sensitivity to rare modifications and specificity for confident functional annotation.

\subsubsection{Validating the Stage 2 Enzyme-Substrate Pairing System (ESPS)}

\textbf{Optimal Fusion Strategy.} To validate the superiority of our dual gated-residual module for fusing enzyme and substrate features, we benchmarked its performance against standard strategies, including concatenation, summation, weighted summation, and a canonical cross-attention mechanism. Our proposed architecture's superior performance across all evaluated metrics (Fig. \ref{fig:ablation}g) validates our gating mechanism, which learns to selectively focus on the most informative features from both the enzyme and substrate to accurately model their interaction.

\textbf{Optimal Input Peptide Length.} To identify the optimal local sequence context for the pairing task, we compared model performance using substrate peptide lengths of 15aa, 21aa, and 51aa, representing a common baseline length, a recent standard utilized by the PTMGPT2 \cite{shrestha2024post} model, and a length consistent with our Stage 1 analysis, respectively. The results clearly indicate that the 15aa peptide length yielded the highest performance, as shown in Fig. \ref{fig:ablation}h. The results clearly indicate that the 15aa peptide length yielded the highest performance (Fig. \ref{fig:ablation}h). This finding suggests that the core determinants of enzyme-substrate specificity are concentrated within the immediate flanking residues, making the 15aa length not only the empirically optimal choice but also the most biologically plausible, as it captures this critical local context while minimizing noise from distal sequences.

\section{Supplementary Methods} \label{secC}

\begin{figure}[htp]
    \centering
\includegraphics[width=1.0 \textwidth]{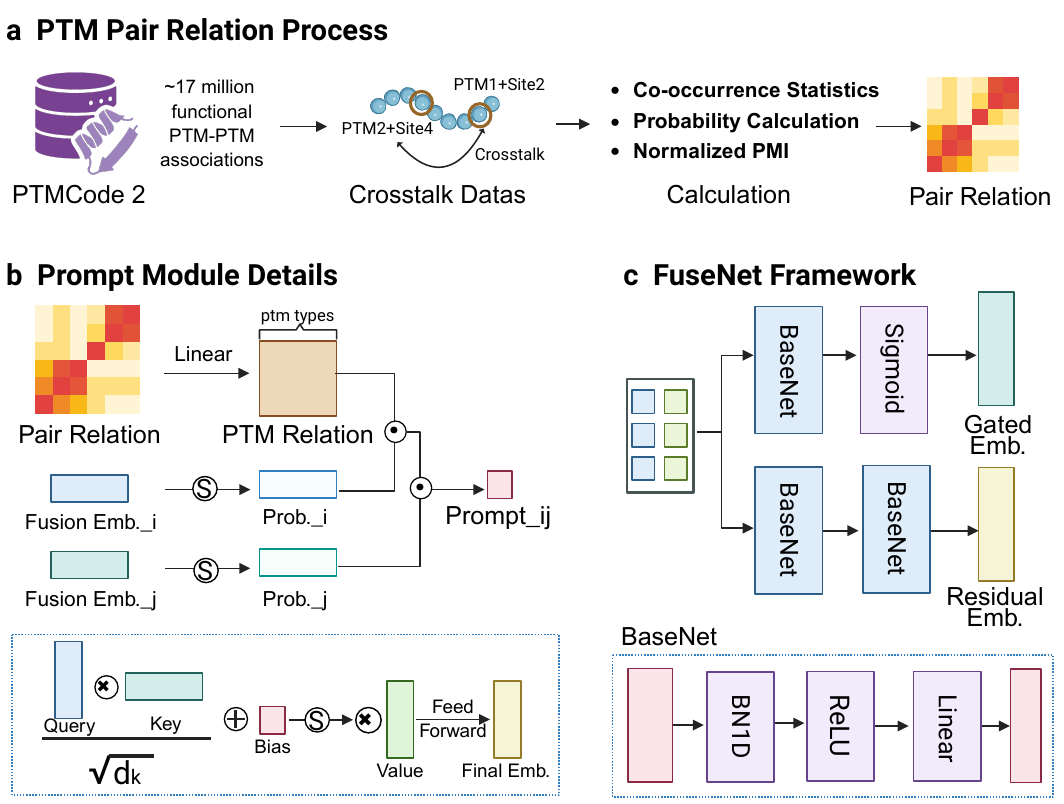}
\caption{\textbf{Submodule of COMPASS-PTM.} \textbf{a}, Creating the PTM relationship map. We process ~17 million PTM interactions from the PTMCode 2 database \cite{minguez2015ptmcode} to build a matrix that quantifies the crosstalk relationships between different PTM types. \textbf{b}, The Prompt Module. This module uses the relationship map from panel \textbf{a} to guide its predictions. It adjusts the score for one PTM based on the likelihood of other PTMs appearing at the same site. \textbf{c}, The FuseNet Framework. This network effectively combines different types of input features using a gating mechanism, which learns to select and prioritize the most important information from each source. The BaseNet is a standard building block used within this network.}
    \label{model_details}
\end{figure}

\subsection{Multi-label Site Profiling Network}
\subsubsection{Implementation of the Fusion Module}\label{secC.stage1fusion}

The Bio-Coupled and Augmented Fusion module (Fig. \ref{model_details}c) operates on the primary protein representation, denoted as $X_p = \mathbf{H}_s^{\text{prot}}$, and the auxiliary chemical representation, $X_a = \mathbf{H}_s^{\text{chem}}$.

\textbf{Gating Mechanism ($\mathbf{\Phi}$). }This component uses the auxiliary features to generate a dynamic gate that modulates the primary features. This process filters out potential noise from $X_a$ and selectively applies useful information. It is formulated as:
\begin{equation}
    \mathbf{\Phi}(X_p, X_a) = \sigma \left(\mathbf{W}_{g2} \cdot \text{ReLU}(\mathbf{W}_{g1} [X_p \,;\, X_a]) \right) \odot X_p
\end{equation}
where $[ \cdot \,;\, \cdot ]$ denotes feature concatenation, $\sigma$ is the sigmoid activation, $\odot$ is element-wise multiplication, and $\mathbf{W}_{g1}, \mathbf{W}_{g2}$ are trainable weight matrices of a two-layer multi-layer perceptron (MLP).

\textbf{Residual Mechanism ($\mathbf{\Psi}$). }This component learns a residual mapping from the concatenated features to capture cross-modal interactions that enrich the final representation. It is defined as:
\begin{equation}
    \mathbf{\Psi}(X_p, X_a) = \mathbf{W}_{r2} \cdot \text{ReLU}(\mathbf{W}_{r1} [X_p \,;\, X_a])
\end{equation}
where $\mathbf{W}_{r1}, \mathbf{W}_{r2}$ are the trainable weights of a separate two-layer MLP.

\subsubsection{Prompting PTM Attention Matrixes} \label{secC.matrix}
The foundation of this prompt is a prior relationship matrix, $\mathbf{R}_{\text{prior}}$, derived from global PTM co-occurrence statistics. We computed this matrix using normalized pointwise mutual information (nPMI) on over 17 million PTM associations from the PTMCode 2 database \cite{minguez2015ptmcode}. The resulting nPMI scores, ranging from -1 (mutually exclusive) to +1 (co-occurring), populate $\mathbf{R}_{\text{prior}}$, which is defined as symmetric with a zero diagonal to isolate crosstalk effects (Fig. \ref{COMPASS-PTM_appendix}c and Fig. \ref{model_details}a).

To adapt the global statistics to task-specific patterns, the static prior $\mathbf{R}_{\text{prior}}$ is transformed into a learnable matrix $\mathbf{R}$ via a projection network $g_\theta$:

\begin{equation}
    \mathbf{R} = g_\theta(\mathbf{R}_{\text{prior}})
\end{equation}

Specifically, $g_\theta$ projects $\mathbf{R}_{\text{prior}}$ using two parallel linear layers, and the resulting matrices are combined via matrix multiplication to produce $\mathbf{R}$. This architecture enables the model to learn complex, non-linear dependencies between PTM types, while initializing the learning process with a strong biological prior (Fig. \ref{model_details}b). 

\subsubsection{Detailed Loss Function Formulation}\label{secC.loss}
\textbf{Macro-Averaged Dice Loss.} The Dice loss ($\mathcal{L}_{\text{Dice}}$) is computed as the macro-average of the Dice coefficient over all $C$ classes. This treats each class equally, regardless of its prevalence. The formulation is:
\begin{equation}
    \mathcal{L}_{\text{Dice}} = 1 - \frac{1}{C} \sum_{c=1}^{C} \frac{2 \sum_{i=1}^{N} y_i^c \hat{y}_i^c + \epsilon}{\sum_{i=1}^{N} y_i^c + \sum_{i=1}^{N} \hat{y}_i^c + \epsilon}
    \label{eq:dice}
\end{equation}
where $N$ is the total number of sites in a batch, $C$ is the number of PTM classes, $y_i^c \in \{0, 1\}$ is the ground-truth label, and $\hat{y}_i^c$ is the predicted sigmoid probability for site $i$ and class $c$. The smoothing factor $\epsilon$ prevents division by zero.

\textbf{Micro Focal Loss.} The Focal loss ($\mathcal{L}_{\text{micro}}$) addresses class imbalance by reducing the loss contribution from well-classified examples, allowing the model to focus on hard negatives and positives. It is defined as:
\begin{equation}
    \mathcal{L}_{\text{micro}} = -\frac{1}{N}\sum_{i=1}^{N} \sum_{c=1}^{C} \left[y_{i}^{c}(1-\hat{p}_{i}^{c})^\gamma \log(\hat{p}_{i}^{c}) + (1-y_{i}^{c})(\hat{p}_{i}^{c})^\gamma \log(1-\hat{p}_{i}^{c})\right]
    \label{eq:focal}
\end{equation}
where $\hat{p}_{i}^{c}$ is the predicted probability for sample $i$ and class $c$. The focusing parameter $\gamma$ (set to 2.0) controls the down-weighting effect.

\subsection{Enzyme–Substrate Pairing System}
\subsubsection{Implementation Details of Dual-Gated Residual Fusion}
\label{secC.stage2fusion}
Our dual-gated architecture symmetrically processes both substrate peptide and enzyme features as co-primary modalities. This is achieved through two reciprocal gating units, where each gate learns to dynamically refine the representation of the opposing modality, thereby removing noise and isolating interaction-specific signals. Concurrently, a residual connection preserves the holistic information from both feature streams, preventing information loss during the filtering process (Fig. \ref{COMPASS-PTM_appendix}d).

\textbf{Feature Integration.} Input features are concatenated as $\mathbf{H}_{\text{combined}} = [\mathbf{H}_{\text{substrate}}; \mathbf{H}_{\text{enzyme}}]$.

\textbf{Gating Networks.} Two independent MLPs generate modality-specific gates:
\begin{align}
    \mathbf{g}_{\text{substrate}} &= \sigma(\text{MLP}_{\text{gate\_sub}}(\mathbf{H}_{\text{combined}})) \\
    \mathbf{g}_{\text{enzyme}} &= \sigma(\text{MLP}_{\text{gate\_enz}}(\mathbf{H}_{\text{combined}}))
\end{align}

\textbf{Residual Composer.} A third MLP learns interaction-specific features:
\begin{equation}
    \mathbf{H}_{\text{residual}} = \text{MLP}_{\text{res}}(\mathbf{H}_{\text{combined}})
\end{equation}



\end{appendices}



\end{document}